| | |
|---|---|
| **Category No.** TN929 | **Public** Yes ■ No ☐ |
| **UDC** 621.39 | **Thesis No.** D-10617-308-2020 |

# Chongqing University of Posts and Telecommunications

# Thesis for Master's Degree

| | |
|---|---|
| **TITLE** | Research on Survivability Strategies of Virtual Network |
| **COLLEGE** | School of Communication and Information Engineering |
| **AUTHOR** | Subhadeep Sahoo |
| **STUDENT NUMBER** | L201720063 |
| **DEGREE CATEGORY** | Master of Engineering |
| **MAJOR** | Information and Communication Engineering |
| **SUPERVISOR** | Ning-Hai Bao, Associate Professor |
| **SUBMISSION DATE** | March 30, 2020 |

# DECLARATION OF ORIGINALITY

I declare that this thesis/dissertation is the result of an independent research I have made under the supervision of my supervisor. It does not contain any published or unpublished works or research results by other individuals or institutions apart from those that have been referenced in the form of references or notes. All individuals and institutions that have made contributions to my research have been acknowledged in the Acknowledgements. I am fully aware that I myself will bear all the legal responsibility arising from violation of the above declaration.

**Signature of Author:** _Subhadeep Sahoo_ **Date:** June 09, 2020

# COPYRIGHT PERMISSION LETTER

I hereby authorize Chongqing University of Posts and Telecommunications (CQUPT) to use my thesis to be submitted to the CQUPT, namely I grant an irrevocable and perpetual license that includes nonexclusive world rights for the reproduction, distribution, and storage of my thesis in both print and electronic formats to CQUPT and to the relevant governmental agencies or institutes to reproduce my thesis. I also extend this authorization to CQUPT, for the purposes of reproducing and distributing single microform, digital, or printed copies of the thesis or dissertation on demand for scholarly uses.

**Signature of Author:** _Subhadeep Sahoo_ **Date:** June 09, 2020

**Signature of Supervisor:** _Ninghai Bao_ **Date:** June 10, 2020

# ABSTRACT

In recent years, the development of diverse range of applications, such as V2V communication, Industry 4.0, medical services, Internet of Things etc., has promoted explosive growth of data traffic and huge consumption of network resources. Massive amounts of devices are integrated into the internet and constantly generating large volumes of data, which need to be converged, processed and stored for a variety of heterogeneous applications. It poses critical challenges to the future networks in terms of network architectures and services. As a promising solution, virtualization is vigorously developed to eliminate the ossification of traditional Internet infrastructure and enhance the flexibility on sharing the substrate network (SN) resources including computing, storage, bandwidth, etc. With network virtualization, cloud service providers can utilize the shared substrate resources to provision virtual networks (VNs) and facilitate wide and diverse range of applications.

As more and more internet applications migrate to the cloud, the resource efficiency and the survivability of VNs, such as single link failure or large-scale disaster survivability, have become crucial issues. The VN embedding problem is non-deterministic polynomial (NP)-hard; so, separating it into node mapping and link mapping sub-problems without any coordination between them might cause a high embedding cost. Provisioning high amount of backup resources to protect against disaster failures can lead to higher blocking ratio of the VNs. Providing protection and/or restoration for VN components with minimal resource consumption without considering the service level agreement (SLA) constraint such as availability of VN applications and service interruptions can cause high downtime of the services. This dissertation presents two independent approaches for distinct scenarios to solve the aforementioned open challenges.

In first approach, we study two-stage coordinated survivable virtual network embedding (SVNE) problem and propose an adaptive path splitting based SVNE (APSS) scheme for EONs, which can combat against single-link failures. In the node mapping stage, APSS exploits an anchor node strategy to balance the node resource utilization and restrict the solution space of the candidate substrate nodes, which can coordinate the node mapping with link mapping to limit the distance spans of the virtual links over the SN. Then, we employ an adaptive path splitting policy to provide full protection against single-link failures with partial backup resource, and design an agile frequency slot windows choosing mechanism to mitigate the spectrum fragmentation for link resource efficiency. We evaluate the proposed APSS scheme and compare its performances with some counterpart schemes. Simulation results demonstrate that the proposed APSS scheme can achieve satisfactory performance in terms of spectrum utilization and blocking ratio, even if with higher backup redundancy ratio.

The second approach studies the VN survivability problem for large-scale disaster scenario and proposes a synchronous evacuation strategy for VNs with dual virtual machines (VMs) inside a disaster risk zone (DRZ), which suffer higher risks than the VNs with single VM inside the DRZ. The evacuation mainly includes two processes, viz., VN reconfiguration and VM live migration. For the threatened VNs, the evacuation strategy first re-maps them at the outside of the DRZ with minimal resource cost, then exploits post-copy technique to sustain the online service alive. During the VM migration process, the operations of basic bandwidth deployment and bandwidth upgradation are implemented, so as to encourage parallel migrations and maximize the resource utilization. The migration strategy also enhances synchronicity among the VMs of the same VN to shorten the dual-VM evacuation time. Numerical results show that the proposed strategy can achieve satisfactory performances in terms of average and total evacuation times of dual-VMs.

So, this dissertation considers two different disaster scenarios and proposes two different strategies to address the survivability issue of the VNs. Finally, this dissertation also briefly introduces our future work, which is VM technology-based service migration strategy for mobile edge computing (MEC) servers to minimize the processing delay and transmission latency for satisfying the quality of services (QoS) of the latency-constrained applications.

# 摘要

近年来，车联网（Vehicle-to-Vehicle，V2V）通信、工业 4.0、医疗服务、物联网等海量异构应用的迅速发展，使得数据流量和网络资源的消耗呈爆炸式增长。大量的终端设备被集成到互联网中，其产生大量的数据需要被聚合、处理和存储，以用于各种异构应用，这对未来的网络体系结构和服务提出了严峻的挑战。虚拟化技术作为一种很有前景的解决方案得到了大力的发展，网络虚拟化可以消除传统互联网基础设施的空间限制，可以增强底层物理网络（Substrate Network，SN）的计算、存储、带宽等网络资源的灵活性。基于网络虚拟化，云服务器提供商可以利用共享的底层资源来提供虚拟网络（Virtual Networks，VNs），以促进广泛多样的应用。

随着越来越多的互联网应用向云端迁移，VNs 的资源利用率和生存性已成为关键问题，如单链路故障或大规模灾难生存性等。VN 映射问题是非确定性多项式（Non-deterministic Polynomial，NP）难问题，因此，将其分为节点映射和链路映射两个子问题，如果两者之间没有任何协调，可能会导致较高的映射代价。为防止灾难性故障而提供大量备份资源可能会导致 VNs 的阻塞率更高。在不考虑服务水平协议（Service Level Agreement，SLA）的约束（如 VN 应用程序的可用性和服务中断）等情况下，以最小的资源消耗为 VN 组件提供保护和/或恢复可能会导致服务的高停机时间。本文针对不同的场景提出了两种独立的方法来解决上述问题。

1.本文研究了两阶段协同生存虚拟网络映射（Survivable Virtual Network Embedding,SVNE）问题,并针对EONs提出了一种自适应的基于路径分割的SVNE（APSS）方案，该方案能够对抗单链路故障。在节点映射阶段，APSS 采用锚节点来平衡节点资源的利用率，并限制候选子节点的解空间，从而协调节点映射和链路映射，限制虚拟链路在 SN 上的距离跨度。然后，采用自适应的路径分割方案，在部分备份资源的情况下，对单链路故障提供充分的保护，并且设计灵活的

时隙窗口选择机制可以减轻频谱碎片对链路资源效率的影响。最后对所提出的 APSS 方案进行了评估，并与其他类似方案进行了性能比较。仿真结果表明，该方案即使具有较高的备份冗余率，也能在频谱利用率和阻塞率方面取得优异的性能。

2.本文通过研究大规模灾难场景下虚拟网络的生存性问题，提出了一种在灾难风险区（Disaster Risk Zone，DRZ）内具有双虚拟机（Virtual Machines，VM）的虚拟网的同步撤离方案，该方案相比 DRZ 内具有单虚拟机的虚拟网络具有更高的风险。撤离方案主要包括 VN 重构和 VM 在线迁移两个过程。对于受到威胁的 VNs，首先在 DRZ 的外部以最小的资源成本对他们进行重新映射，然后利用后复制技术来维持在线服务。在虚拟机迁移过程中实现了基础带宽配置和带宽升级操作，以鼓励并行迁移，最大化资源利用率。该方案还增强了同一 VN 的 VM 之间的同步性，缩短了双 VM 的撤离时间。仿真结果表明，所提出的方案在平均撤离时间和总撤离时间上均能达到令人满意的效果。

综上所述，本文基于不同的灾难场景，提出了两种不同的方案来解决 VNs 的生存性问题。最后，本文对相关研究进行了展望，对移动边缘计算（Mobile Edge Computing，MEC）服务器，基于 VM 迁移技术的服务迁移方案来最小化处理时延和传输时延，从而满足时延受限应用的服务质量（Quality of Services，QoS）。

**关键词：**灾难风险，弹性光网络，频隙，在线迁移，生存性，同步撤离，虚拟机，虚拟网映射

# CONTENTS

# List of Figures

# List of Tables

# Abberviations

| | |
|---|---|
| ABR | Average Backup Redundancy-ratio |
| AET | Average Evacuation Time |
| AGC | Average Guard Band Consumption |
| ANS | Anchor Node Strategy |
| APS | Adaptive Path Splitting |
| APSS | Adaptive Path Splitting based SVNE |
| ASC | Average Spectrum Consumption |
| BEDV | Best-effort Evacuation for Dual VMs |
| CPU | Central Processing unit |
| DCM | Dispersion Compensation Module |
| DPP | Dedicated Path Protection |
| DRZ | Disaster Risk Zone |
| EON | Elastic Optical Network |
| FCM | FSWs Choosing Mechanism |
| FF | First Fit |
| FSW | Frequency Slots Window |
| GRC | Global Resource Capacity |
| InP | Infrastructure Providers |
| IoT | Internet of Things |
| ISP | Internet Service Providers |
| ITU-T | International Telecommunication Union Telecommunication |
| LAG | Layered Auxiliary Graph |
| LL | Link List |
| MEC | Mobile Edge Computing |
| NP | Nondeterministic Polynomial |
| NSFNET | National Science Foundation Network |
| OFDM | Orthogonal Frequency Division Multiplexing |
| PDPP | Partitioning Dedicated Path Protection |
| QAM | Quadrature Amplitude Modulation |
| QoS | Quality of Service |
| QPSK | Quadrature Phase Shift Keying |
| RSA | Routing and Spectrum Assignment |
| SBVT | Sliceable Bit Rate Variable Transponders |
| SEDV | Synchronous Evacuation Strategy for Dual VMs |
| SLA | Service Level Agreement |
| SN | Substrate Network |
| SP | Service Providers |
| SRG | Shared Risk Group |
| SVNE | Survivable Virtual Network Embedding |
| TET | Total Evacuation Time |
| UE | User |
| USNET | United State Network |
| V2V | Vehicle to Vehicle |
| VBR | VN Request Blocking Ratio |
| VM | Virtual Machine |
| VN | Virtual Network |
| VNE | Virtual Network Embedding |
| WAN | Wide Area Network |
| WDM | Wavelength Division Multiplexing |

# Chapter 1 Introduction

## 1.1 Optical Substrate Networks

Optical fibers are increasingly deployed in fiber-optic communication, as they allow transmission over longer distances and at higher bandwidths (data rates) than wire cables. Both properties are due to lower losses and greater number of channels that can be transmitted across their broad-spectrum range simultaneously [1].

The region between 1.3 and 1.6 μm is used for transmission in optical fibre. Within this region, the C band exhibits the lowest losses of the entire fiber spectrum and is utilized for transmission over very long distances (from tens to thousands of kilometres). The C band refers to the wavelengths around 1550 nm and includes wavelengths between approximately 1525 nm (or a frequency of 195.9 THz) and 1565 nm (191.5 THz) [2].

### 1.1.1 Traditional-Wavelength Division Multiplexing Optical Networks

Wavelength Division Multiplexing (WDM) denotes the technology which enables a number of optical signal carriers to be transmitted to a single optical fiber utilizing different wavelengths. The International Telecommunication Union Telecommunication (ITU-T) Standardisation Sector has specified a table of all the wavelengths (and their corresponding central frequencies) to enable unified usage of the entire C band; this list is outlined in recommendation ITU-T G.694.1[3]. Initially, WDM wavelengths were located in a grid of precisely 100 GHz (about 0.8 nm) of optical frequency range, with a reference frequency fxed at 193.10 THz (1552.52 nm).

In the last decade, a large number of significant innovations have increased capacity by a factor of around 20 (compared to legacy WDM systems at 10 Gb/s on a 100-GHz spacing) to confront the constant growth in traffic. First, it was possible to squeeze channels by spacing them 50 GHz apart (about 0.4 nm) in core networks, enabling about 80 (and most recently 96) channels to be transported [4]. Moreover, the channel bandwidth has been improved by building optoelectronic equipment at higher bit levels, often working inside the 50 GHz system, going from 2.5 to 10 Gb/s, then to 40 Gb/s, and up to 100 Gb/ per wavelength since 2010; these channel width and bit rate enables 2 Bit/s/Hz spectral performance [5].

Unlike the early days of WDM networks, when an optical fiber bandwidth was known to be infnite, the optical spectrum would be a finite resource in the immediate future and nowadays industry is researching ways to increase the overall spectrum performance. Improvements in signal transmission techniques have enabled the spectrum occupied by the optical signals to be reduced: the combination of coherent detection techniques with Nyquist pulse shaping allows 100 Gb/s to be transmitted in as little as 33 GHz [2].

Increasing the single carrier bit rate over 100 Gb/s involves the use of higher order Quadrature Amplitude Modulation (QAM)- for e.g., 16-QAM doubles the bit rate in comparison with Quadrature Phase Shift Keying (QPSK) and therefore, offers 200 Gb/s capacity. However, those higher QAM formats operate only over shorter transmission lengths. One means of integrating higher-resolution channels like 400 Gb/s and 1 Tb/s is by introducing multi-carrier signals at at least the same symbol rate as 100 Gb/s. For example, a 400 Gb/s channel can be obtained with two 16-QAM modulated subcarriers (200 Gb/s each), each at 37.5 GHz, for a total bandwidth of 75 GHz; and a 1 Tb/s channel can be obtained with four 32-QAM modulated subcarriers, each at 43.75 GHz, for a total bandwidth of 175GHz, respectively [6]. The first consequence in this application is the need for greater channel size, which violates the traditional 50 GHz grid per channel regulation.

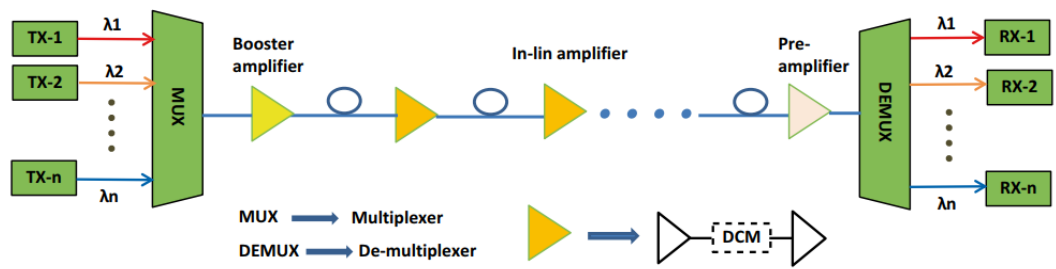

Fig. 1.1. Point-to-point WDM system.

Fig. 1.1 illustrates a point-to-point WDM system schematic in which several wavelength channels are created in optical transmitters, each of which is modulated by a data signal and then combined by a WDM multiplexer. The composite WDM signal is then distributed via an optical fiber connection with optical amplifers to enhance the signal before delivery, to account for the fiber degradation at each span, and to increase the sensitivity of the receiver. As shown in the inset in Fig. 1.1., the in-line amplifers are normally two stage amplifers, where the Dispersion Compensation Module (DCM) is used for the correction of chromatic fiber dispersion in each span. Alternatively, the fiber dispersion can be balanced in the coherent receivers by utilizing coherent equipment. Therefore, the latest WDM networks are planned to be DCM-less to allow optimal efficiency with coherent infrastructure for higher speeds. A WDM de-multiplexer is used at the transmitting end to split the WDM signal into different channels, and the optical receivers retrieve data signals [2, 6].

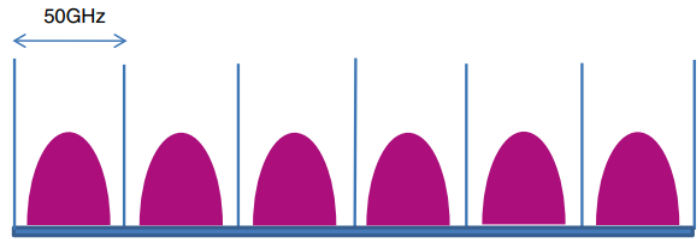

Fig. 1.2. Fixed 50GHz ITU grid (each grid is 200 Gb/s, DP-16QAM).

WDM technology is an efficient way to increase network capacities. Although bandwidth is improved by the amount of WDM channels, all of the WDM channels use the same network infrastructure, including optical fibre, optical amplifiers, DCMs, WDM multiplexer, and de-multiplexer. Thus, the network cost is also shared, resulting in lower cost per channel. If traffic demand rises, network capacity will improve with the introduction of extra transponders (i.e. transmitters and receivers) at the traffic termination points. Additionally, the WDM system is open to the wavelength channels holding data signals. Different data speeds and modulation types will also be added to help improve network capacity [2].

Table 1.1. Total fiber capacity using WDM in C band with 50 GHz grid.

| Data rate (Gb/s) | Total capacity (Tb/s) | | | Spectral effciency (bits/s/Hz) |
|---|---|---|---|---|
| | 80 λs | 88 λs | 96 λs | |
| 2.5 | 0.2 | 0.22 | 0.24 | 0.05 |
| 10 | 0.8 | 0.88 | 0.96 | 0.2 |
| 40 | 3.2 | 3.52 | 3.84 | 0.8 |
| 100 | 8 | 8.8 | 9.6 | 2 |

For instance, when implementation of the WDM system started in mid-1990, the prevailing data rate was 2.5 Gb/s. The data rate increased to 10 Gb/s as technologies grew, and then to 40 Gb/s utilizing different modern modulation formats. More recently, 100 Gb/s has been mainstreamed and readily available, with developments in coherent technologies and optical signal processing. This is already being implemented quickly throughout the networks. Table 1.1 displays the overall power of fibers in a 50 GHz grid using WDM. For industrial WDM networks, over the years, gross fiber bandwidth has grown from up to 0.24 Tb/s at a data rate of 2.5 Gb/s to a peak of 9.6 Tb/s at a data rate of 100 Gb/s higher [6]. It should be remembered that the usage of optical fiber has also improved dramatically, as shown by increasing the spectral quality from 0.05 to 2 bit/s/Hz as the data rate increases to 100 Gb/s. The spectral efficiency tends to grow when the data speed rises above 100 Gb/s and as the network infrastructure transitions from fixed-grid WDM to scalable grid networking, which will be discussed in more depth in the sub-section below.

## 1.1.2 Emerging-Elastic Optical Networks

For the current networks, it became necessary to have a new standardized grid allowing the best spectral effciency for the increased diversity of the spectrum requirements (e.g. a 100 Gb/s channel ftting in 37.5 GHz and 400 Gb/s in 75 GHz) [7].

To resolve this problem, the ITU-T introduced a fner grid associating a variable frequency slot with an optical link, called a flexible frequency grid or, more generally, a flex-grid. This Flex-grid optical network is also known as Elastic Optical Network (EON). Flex-grid enables a specific number (n) of fxed-sized slots to be assigned to an optical channel according to the requirements. A slot weighs 12.5 GHz, enabling 100 Gb/s channels to be broadcast in 37.5 GHz (n=3), instead of 50 GHz in the fxed-grid. The effect of flex-grid on transmission is shown in Fig. 1.3. In this example, flex-grid can support 8 rather than 6 200 Gb/s channels, but it can also group these into super-channels, if needed, which can be transmitted as a single entity via an optical network [2].

Since the introduction of emerging technology, communication networks are evolving rapidly. Flex-grid enables partitioning of the optical spectrum with a great deal of fner granularity, and randomly large spectral slots may even be set up and distributed around a network. Flex rate allows transceivers to operate for various modulation formats in order to obtain the maximum spectral efficiency for specified optical paths. Flex-grid includes novel LCoS technologies to enable fne optical filter control, while flexrate utilizes recent advances in coherent transmission to include variations in the modulation format. Combining these two definitions allows for the design of broad bandwidth cables, in which several sub-channels are placed together to create a super-channel, providing bandwidths in the Tb/s scale. This is a bold new approach for key communication networks, offering extremely scalable future optical transport that utilizes the existing optical fibre spectrum resources to their maximum capacity [8].

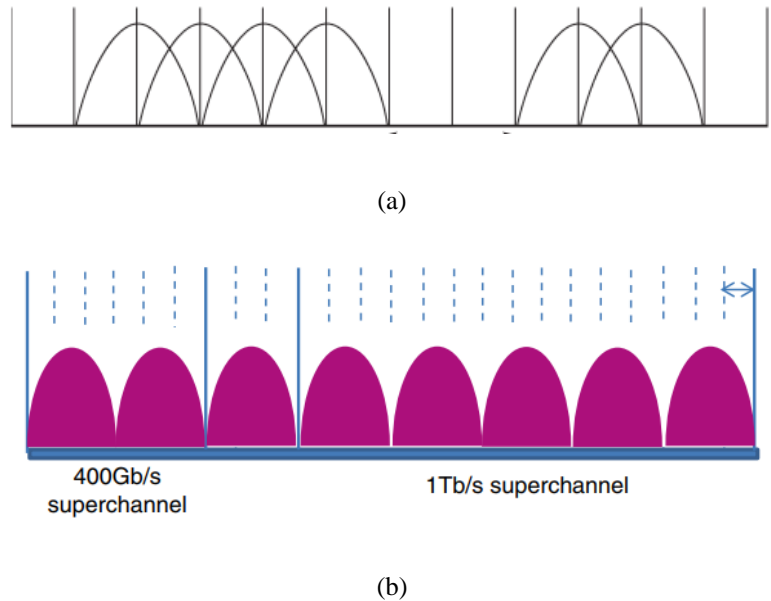

(a)

(b)

Fig. 1.3. (a) Overlapping subcarriers caused by OFDM technology in EON (b) 12.5 GHz resolution for flex-grid allows closer packing of channels.

This flexibility in optical networking would enable metro-core networks to become more robust and theoretically flatter in the sense that as the network expands and develops, new direct connections between nodes will be feasible. Ultimately, core networks must develop from rigid fixed optical spectrum pipes to dynamically flexible resources, capable of placing arbitrary amounts of spectrum services between nodes on demand and carrying several Tb/s of data [9]. Such improvements would be as revolutionary as the initial WDM idea which will enable our core networks to continue to meet increasingly growing demands on bandwidth for the near future. Despite this, lack of ability isn't the only impetus behind migration. There are other drivers that justify the gradual deployment of flex-grid technology, which are as follows.

- The usage of higher data rates and specialized modulation formats for different links in the short term (2014-2016) would enable for cost-effective 400 Gb/s (and beyond) signals.

- The introduction of commercial Sliceable Bit Rate Variable Transponders (SBVT) will be a crucial event in the medium term (2017-2019) to expand the coverage of flex-grid regions to certain sections where, while bandwidth is not yet saturated, the opportunity to split several flows would be helpful. SBVTs are BVTs with an integration level that requires several modulators: they are therefore able to build either wide super-channels or smaller individual channels when needed.

- In the long term (>2020), resulting from grappling with projected traffic volumes of hundreds of Tb/s or even several Pb/s that entail the possibility of implementing flex-grid of various network architectures throughout the network. Legacy fixed-grid infrastructure will then be completely converted to core flexgrid.

Fig. 1.4. illustrates the feature of flexible spectrum allocation according to client's requirement in EONs. EON's benefits over WDM include network segmentation, network compression, effective accommodation of several data levels, dynamic distributed resource variability, reach-adaptable line rate, etc. [10].

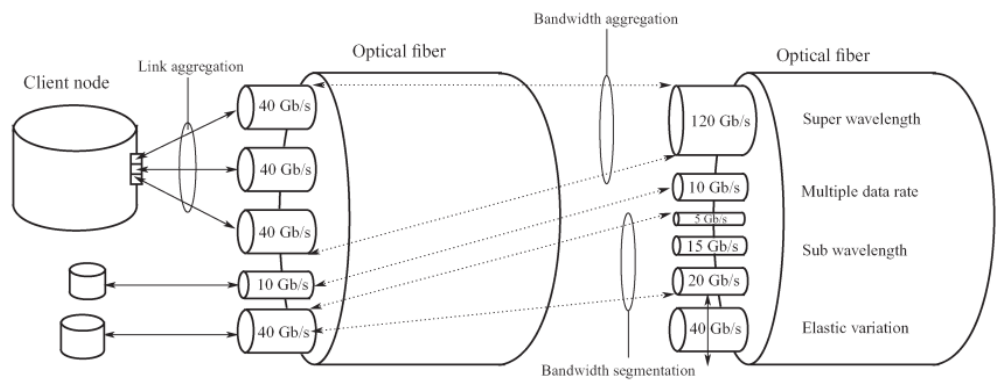

Fig. 1.4. Unique characteristics, namely- bandwidth segmentation, bandwidth aggregation, accommodation of multiple data rates, and elastic variation of allocated resources, of EONs.

- **Bandwidth segmentation:** Traditional optical networks demand that wavelength bandwidth be entirely assigned to an optical path from an end-node pair. Nevertheless, elastic optical networks have a spectrum-efficient segmentation of bandwidth (sometimes named subwavelength) system that delivers a fractional access capacity to the bandwidth. If only partial bandwidth is needed, as seen in

Fig. 1.4., elastic optical network can assign only enough optical bandwidth to satisfy customer traffic, where an optical bandwidth of 40 Gb / s is split into three subwavelengths, such as —5 Gb/s, 15 Gb/s and 20 Gb/s. At the same time, each node on the optical path route allocates a cross-connection with the required spectrum bandwidth to establish an optical end-to-end path of an acceptable distance. The optimal usage of network capital would require fractional bandwidth access to be delivered cost-effectively.

- **Bandwidth aggregation:** Link aggregation is an IEEE 802.3 unified packet-networking system. It integrates several physical ports/links within a switch/router into a single logical port/link to enable exponential growth in link speed as traffic demand grows outside the limits of any single port/link. Likewise, the elastic optical network allows the aggregation function of the bandwidth and hence can establish an optical path of super-wavelength contiguously merged within the optical domain, thereby ensuring a strong use of spectral resources. This particular attribute is seen in Fig. 1.4., where three 40 Gb/s optical bandwidths are multiplexed by an optical Orthogonal Frequency Division Multiplexing (OFDM), to include a 120 Gb/s super-channel [11].

- **Efficient accommodation of multiple data rates:** As shown in Fig. 1.4, the elastic optical network has the ability to provide the spectrally-efficient direct accommodation of mixed data bit rates in the optical domain due to its flexible spectrum assignment. Traditional optical networks with fixed experience the loss of optical spectrum because of the disproportionate frequency distribution for weak bit rate signals.

- **Reach-adaptable linerate:** The elastic optical network has the potential to accommodate realistic line rate as well as simultaneous expansion and contraction in bandwidth by modifying the amount in subcarriers and modulation types.

- **Energy saving:** This facilitates energy-efficient operations to reduce resource usage by disconnecting some of the OFDM subcarriers when traffic is slack.

- **Network virtualization:** It enables virtualization of the optical network with virtual connections assisted by subcarriers OFDM.

## 1.2 Virtual NetworkTechnologies

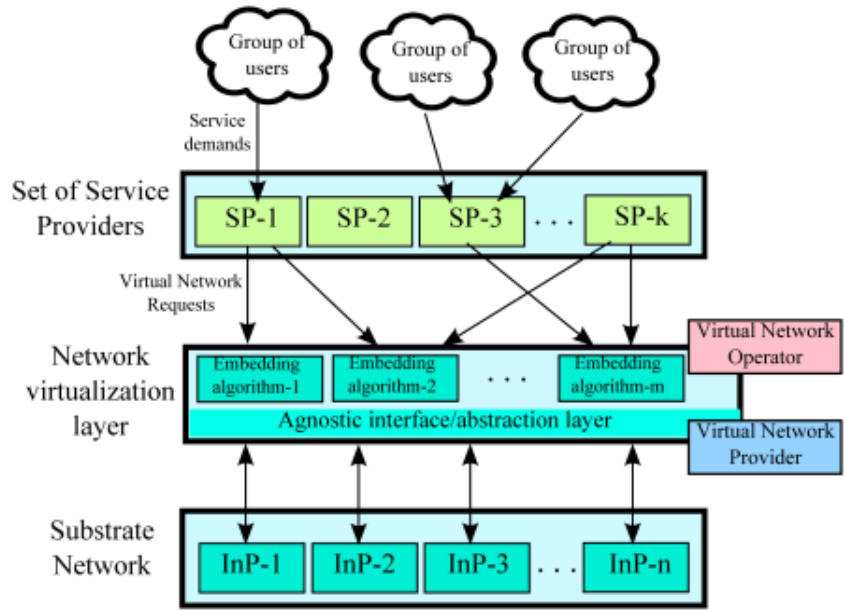

Fig. 1.5. Network vitualization.

In recent years, virtualization overcomes the ossification of traditional Internet and enables great flexibility on sharing the common physical infrastructure such as computing resource and transmission bandwidth, which significantly encourages the development of wide and diverse range of cloud services and applications [12]. Under the network virtualization model, internet service providers (ISPs) can be decoupled into two entities viz., infrastructure providers (InPs) and service providers (SPs). As illustrated in Fig. 1.5., the InP is in charge of deploying and managing physical network infrastructure, while the SP can rent these substrate network (SN) resources from InPs to provision virtual networks (VNs) and serve end users with heterogeneous VN applications [13, 14].

### 1.2.1 Virtual Network Construction and Mapping / Embedding

In general, a VN is composed of several virtual nodes interconnected by virtual links. The VN provisioning problem is also known as VN mapping/embedding (VNE). VNE can be divided into two sub-parts as follows [15, 16]. Fig.1.6. illustrates the VNE model, in which the VNs from SPs are embedded on the SN according to the above-mentioned constraints.

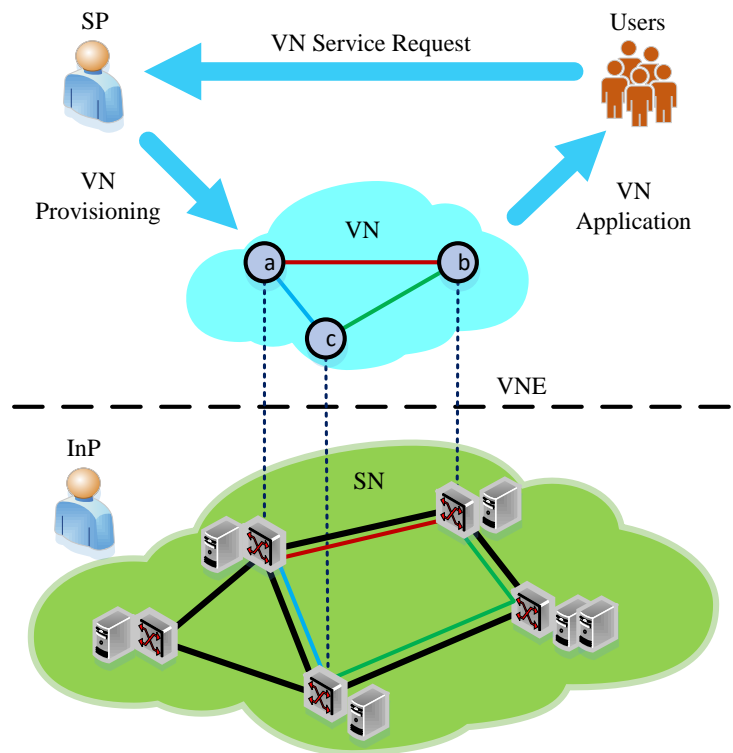

Fig. 1.6. Virtual network embedding model.

- **Node mapping:** Virtual nodes can be realized as the virtual machine (VM) running on a physical server (substrate node). Embedding the virtual nodes on substrate nodes must follows the three constraints.
    (1) The candidate substrate node should have adequate computing (and storage) resource to satisfy the requirement by the virtual node.
    (2) A substrate node can hold only one virtual node from the same VN.
    (3) A virtual node can be mapped only on a single substrate node.
- **Link mapping:** Deploying one or several successive substrate links to establish a virtual link with demanded bandwidth for a corresponding virtual node pair.

### 1.2.2 Virtual Machine Migration

Cloud servers or data centers utilize virtualization for optimized resource control to reduce cloud storage expenses and the allocation for resources. Virtualization allowed by virtual machine (VM) relocation satisfies the ever-increasing demands of complex workload by relocating VMs to another server or data denter. VM migration aims to meet complex resource management targets such as load balancing, power control, fault tolerance, and device maintenance. A live VM migration over a wide area network (WAN) involves transferring CPU state, memory pages, and disk storage from source server to geographically distant destination server without interrupting online services [17, 18]. It facilitates the following features:

- **Load Balancing:** It also needs simultaneous migration of VM(s) when the load is considerably unbalanced and experiences service downtime. It is used for continuing services after fail-over of components which are monitored continuously then load on host distributed to other hosts and no longer sends traffic to that host.

- **Proactive fault tolerance:** Fault is another challenge to guarantee the critical service availability and reliability. Failures should be anticipated and proactively handled, to minimize failure impacts on the application execution and system performance. For this different type of fault tolerance techniques are used.

- **Power management:** Switch the idle mode server to either sleep mode or off mode depending on resource requirements, resulting in substantial energy savings as idle mode server uses 70 percent of its peak power, so consolidating the operating VM to fewer active hosts contributes to major energy savings. So dynamic allocation of VM's to few active servers as much as possible, VM live migration is a good technique for cloud power efficiency.

- **Resource sharing:** The sharing of available hardware resources such as memory, cache, and Processor cycles contributes to deterioration of the application output. Relocating VM's from overloaded server to underloaded server will solve this issue. While, resource sharing contributes to reduced running costs due to switching off the redundant or idle servers.

- **Online system maintenance:** A specific network that requires to be updated and maintained, then all of the specific server's VMs ought to be transferred to an alternative support and operation site the consumers can reach without interruption.

The fundamental strategies of live VM migration can be categorized into pre-copy and post-copy [19].

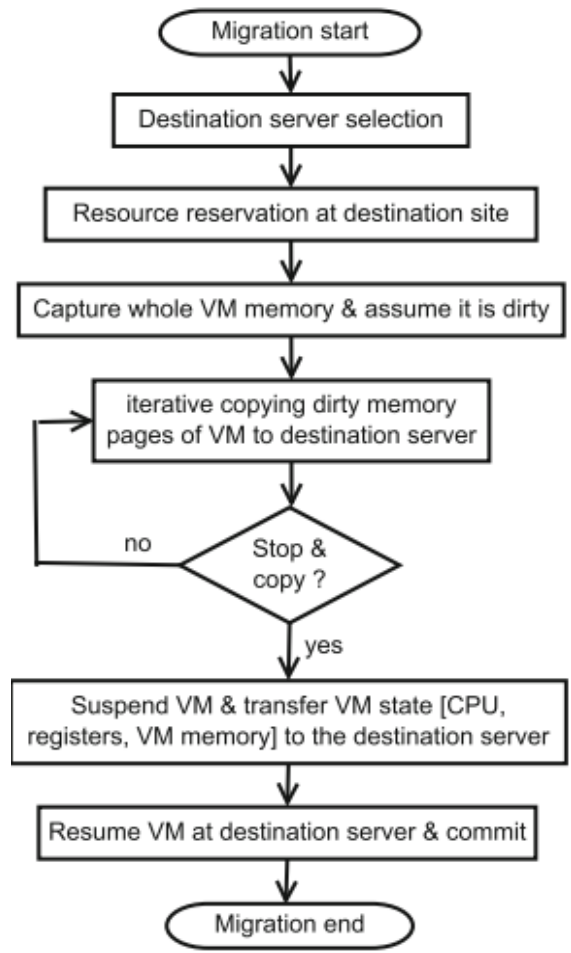

Fig. 1.7. Pre-copy VM live migration.

For pre-copy, it first copies all memory pages and disk storage to the destination server, while the VM is still running on the source server and generating dirty pages (modified memory pages). Then, these dirty pages have to be re-copied iteratively to the destination at a rate higher than the dirtying rate. Finally, the source VM is suspended (down), the CPU state as well as the remaining dirty pages are transferred, and the destination VM is resumed. Most of the studies regarding to pre-copy technique [20], concentrated on the trade-off between the down time and the total migration time by optimizing the pre-copy iterations. Fig. 1.7. illustrates the flowchart of pre-copy live migration.

In contrast, post-copy first suspends the source VM, transfers its CPU state to the destination server, and resumes the destination VM, then transfers memory pages and disk storage to the destination server. As the source VM does not generate dirty pages, the memory pages and disk storage are actively pushed to the destination server only once. The existing studies regarding to post-copy technique usually focused on optimizing the total amount of data to be transferred by minimizing the duplicated memory pages and avoiding page faults [21]. Fig. 1.8. illustrates the flowchart of post-copy live migration.

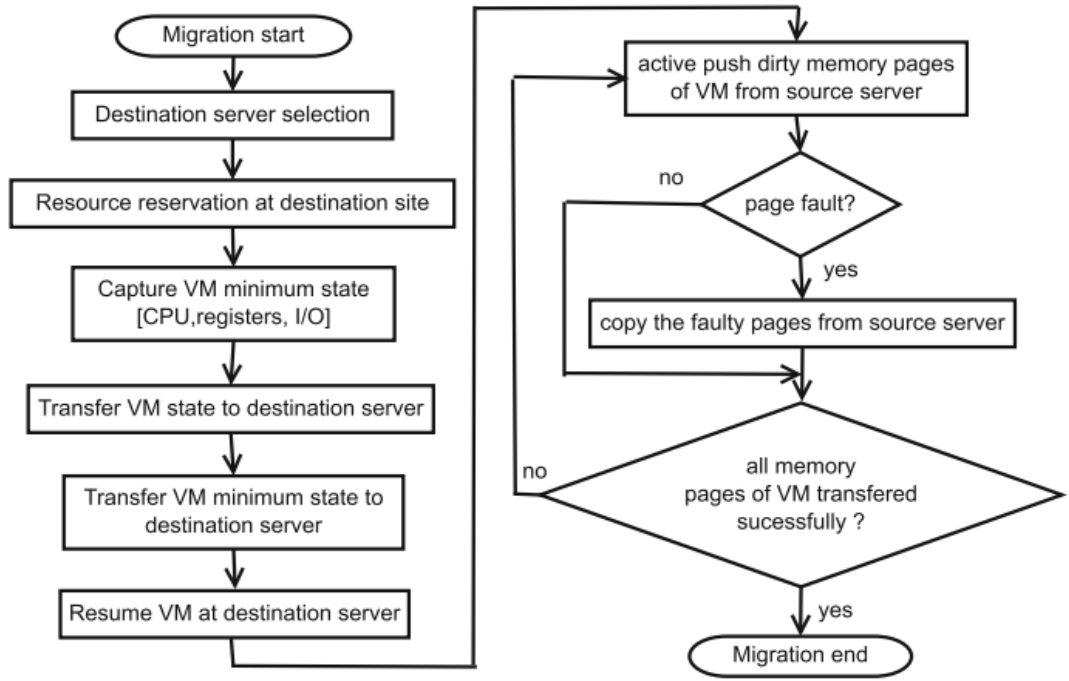

Fig. 1.8. Post-copy VM live migration.

Apparently, the pre-copy is good at reliability, as it can restore the current state of the VM at the source even if the migration fails. But the post-copy is featured for short and predictable migration time because of no dirty pages generated and transferred iteratively.

## 1.3 Virtual Network Survivability

As more and more heterogeneous applications migrate to the cloud, the survivability of VNs has become a crucial issue. With huge transmission capacities, the substrate network (e.g., EONs) can accommodate enormous amount of data and services. So, a substrate link failure, such as an optical fiber cut, might simultaneously damage multiple VNs and disrupt a lot of VN applications. Particularly, during large-scale natural disasters (e.g., earthquake, tsunami, hurricane, etc.), some substrate nodes might be impacted or destroyed, even if the node equipment usually has higher anti-destruction grade than the long spanning cables. It would cause many VM failures and a large amount of VN service losses. Therefore, survivability is a crucial challenge of the VNs.

Protection and restoration are the two traditional strategies for VN survivability. Many literatures researched VN protection schemes, i.e., proactively provisioning backup virtual nodes and/or virtual links during VN embedding (VNE) phase, which is also known as survivable VNE (SVNE).

## 1.3.1 Protection Strategy

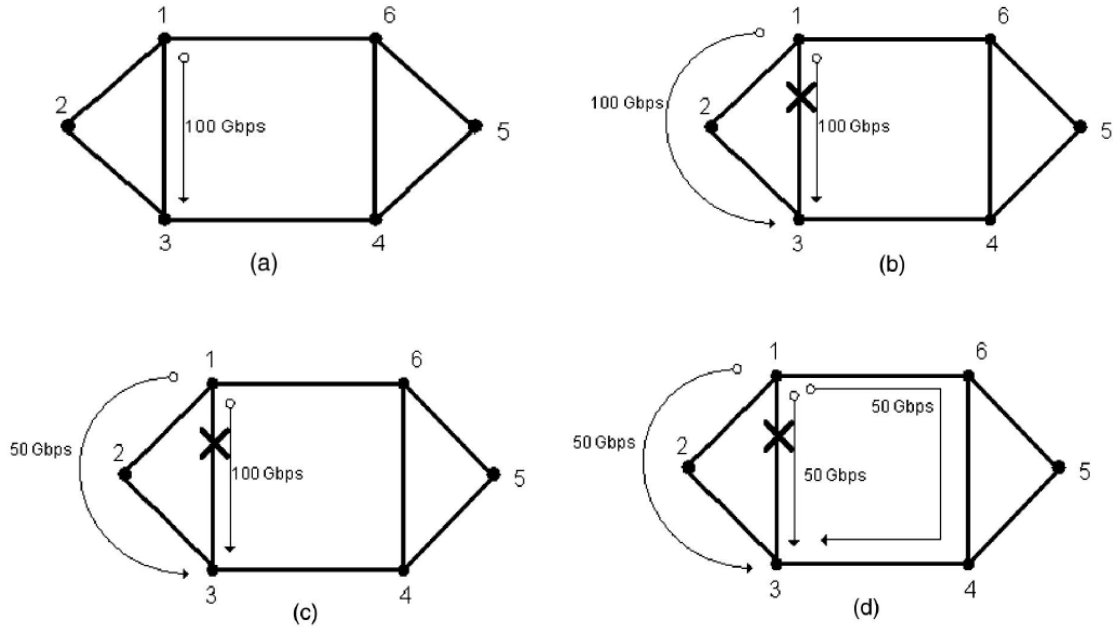

Fig. 1.9. Protection mechanism examples for a source–destination traffic transmission: (a) No protection, (b) Dedicated protection, (c) Dedicated partial protection, and (d) Multipath protection.

In Fig. 1.9, the survivability of a virtual link is explained. For example, a virtual link with the bandwidth requirement of 100 Gbps is mapped between substrate nodes 1 and 3. In Fig. 1.9 (a), there is only one working path without any backup path. So, if the path 1-3 is failed, then the virtual link will also be failed. But as in Fig. 1.9 (b), along with the working path 1-3, there is a backup path 1-2-3 with full demanded bandwidth. So, if path 1-3 is failed, then path 1-2-3 will be activated and the connection will not ne lost. However, providing full bandwidth on the backup path consumes much bandwidth and it costs high. So, we illustrate partial backup in Fig. 1.9 (c). But as in Fig. 1.9 (d), the working path is divided into two sub-paths 1-3, and 1-6-4-3 with 50 Gbps on each sub-path. So, we need only 50 Gbps on the backup path 1-2-3 to support the full protection. This approach can save the bandwidth as well as the cost.

The authors in [22] investigated the SVNE problem to combat against single substrate node failures, in which isomorphism graph, multi-commodity network flow, and VN level backup-sharing are considered respectively to improve resource efficiency. In [23], the authors proposed a probabilistic protection scheme for multiple substrate node failures, in which backup resource sharing is facilitated among different VNs to minimize the consumption of backup computing capacity. In [24], the authors designed location-constrained SVNE schemes with bipartite graph matching and resource sharing policies to protect against single substrate node or link failures. And the author in [25] investigated 1+1 survivable VN mapping schemes to provide full protection with dedicated backup VNs. In a regional failure scenario, literature [26] studied virtual node protection scheme with the consideration of geographical isolation and failure dependent protection. Based on continuous and acknowledged replication of VMs, literature [27] designed a backup VM placement scheme against disaster failures with low transmission latency and minimal node resource consumption. As protection is usually resource-intensive and costly, particularly, some natural disasters (e.g., earthquake and tsunami) are difficult to predict, VN protection against disaster failures might be somewhat scattershot and lack of resource efficiency.

### 1.3.2 Restoration Strategy

VN restoration, on the other hand, can achieve better adaptability and resource efficiency in various failure scenarios. Literatures [28-32] researched VN restoration schemes, i.e., reactively re-provisioning the VN (or part of the VN) for surviving against failures. The authors in [28] designed a VN re-embedding algorithm to recover VNs from SN failures with the objective of minimizing re-embedding cost. In [29], the authors combined the preventive and reactive approaches together and proposed a multipath embedding and opportunistically recovering based survivable scheme to mitigate the impact of malicious attacks. In [30], the authors focused on the problem of recovering VNs from a single substrate node failure and designed a generalized recovery approach to re-embed the affected VNs with resource optimization objectives. In [31], the authors studied progressive recovery schemes against disaster failures, which optimize the residual substrate resource to improve the VN restoration ratio. And the authors in [32] studied multicast VN restoration with the end to end delay constraint for multiple substrate node and/or link failures.

### 1.3.3 VN Survivability with VM migration Strategy

Besides the aforementioned VN protection and restoration schemes, the VM live migration can also be another solution for some survivability scenarios, e.g., combating against potential disaster failures. With which, the threatened VMs can be migrated to safe locations without interrupting the online services. As VM live migration enable continuous service provisioning during the migration process, it can be implemented to sustain the online services alive during a disaster. Thus, migrating as many VMs as possible out of the disaster risk zone (DRZ) can be a feasible solution for VN disaster survivability.

## 1.4 Open Challenges and Major Contributions

The survivability is a crucial challenge of the VNE, which is induced to survivable VNE (SVNE) problem. For this problem, a major challenge is to maximize the acceptance ratio of VN requests with minimum cost of physical resources, while the bandwidth resource is usually a critical bottleneck of SNs. Also, most of the existing survivability works mainly focused on providing protection and/or restoration for VN components with minimal resource consumption, but neglected the service level agreement (SLA) constraint such as availability of VN applications and service interruptions.

In this dissertation, we research on the SVNE problem for EONs and propose an adaptive path splitting based SVNE (APSS) scheme. Our main concerns are the acceptance ability of VN requests and the utilization of spectrum resource on EONs, while guaranteeing the full survivability against single-link failures. Simulation results demonstrate that APSS scheme outperforms baseline schemes in terms of blocking ratio and resource efficiency. In chapter 3, in order to survive the disaster-threatened VNs and sustain their online services, we propose synchronous evacuation scheme to evacuate VNs out of the DRZs. The main objective of this scheme is to minimize the evacuation time period, as well as the entire time period for completing all the evacuations.

## 1.5 Thesis Organization

The remainder of the thesis is organized in the following fashion. Chapter 2 discusses the Adaptive Path Splitting Based SVNE in EONs. In this chapter, we discuss protection scheme for VN survivability issue. Then we propose a scheme named as APSS to provide full protection against single-link failures and achieve satisfactory performance in terms of spectrum utilization and blocking ratio. In chapter 3, we discuss Synchronous Evacuation of VMs Under Disaster Risks. Here, we discuss the reactive survivability scheme and evacuate the VMs from the disaster threatened region to the safer locations with an objective of minimum evacuation time. We briefly discuss our future work on VM based service migration strategy at MEC in chapter 4. Finally, the dissertation is concluded in chapter 5.

# Chapter 2 Adaptive Path Splitting Based SVNE in EONs

## 2.1 Chapter Overview

Network virtualization facilitates heterogeneous cloud applications to share the same physical infrastructure with admirable flexibility, while resource efficiency and survivability are critical concerns for VNE. In this chapter, we study two-stage coordinated SVNE problem and propose an APSS scheme for EONs. For APSS, we first develop a concise anchor node strategy to balance the node resource utilization and restrict the solution space of the candidate substrate nodes, which coordinates node mapping with link mapping to limit the distance spans of the virtual links. Then, we employ an adaptive path splitting policy to provide full protection against single-link failures with partial backup resource, and design an agile frequency slot windows choosing mechanism to mitigate the spectrum fragmentation for link resource efficiency. We evaluate the proposed APSS scheme and compare its performances with some counterpart schemes. Simulation results demonstrate that the proposed APSS scheme can achieve satisfactory performance in terms of spectrum utilization and blocking ratio, even if with higher backup redundancy ratio.

The major contributions of this chapter are as follows.

- We propose an anchor node strategy for the node mapping stage. In which, a substrate node with the highest available capacity of node resource is defined as the anchor node for the virtual node with the highest demanded capacity of the node resource. Then a distance parameter is set to restrict the solution space of the candidate substrate nodes for the rest of the virtual nodes. In this way, it can balance the node resource utilization and coordinate the node mapping with link mapping to limit the distance spans of the virtual links over the SN.

- For single-link-failure protection deployment, the fixed path splitting, i.e., evenly splitting the demanded bandwidth over a given number of disjoint paths, can significantly reduce the backup redundancy, but there are still several concerns aforementioned need to be solved. In this chapter, we introduce an adaptive path splitting policy in link mapping stage, in which the number of splitting paths and the spectrum resource assigned on each path is adaptively determined according to the SN status. Simulation results demonstrate that the adaptive path splitting, even

        if with high backup redundancy, can dramatically outperform the fixed path splitting in terms of average spectrum consumption.
- In order to improve the link resource utilization during link mapping stage, we propose a FS windows choosing mechanism to solve the spectrum fragmentation problem. In which, a cost function is designed to evaluate the costs of the candidate FS windows (FSWs) along a substrate path, which can help to find an appropriate FSW for spectrum allocation while saving more continuous and contiguous FSs in the EON for the upcoming VN requests.

The rest of the chapter is organized as follows. In section 2.2, the background and related work are summarized. The motivation of this chapter is introduced in section 2.3. Section 2.4 presents the network model and problem statement. In section 2.5, the detailed APSS scheme is proposed and the corresponding heuristic algorithms are developed in section 2.6. Finally, section 2.7 summarizes this chapter.

## 2.2 Background and Related Work

The SVNE problem has been widely investigated in existing literatures [33, 34, 25, 35-43]. To guarantee the quality of service (QoS) of VN applications even after the failures, proactive provisioning of backup resources during the embedding phase is necessitated and it is named as VN protection. In [25, 35], the authors studied 1+1 SVNE schemes to provide full protection with dedicated backup VNs. Literature [25] proposed a dedicated VN protection scheme, in which all the substrate nodes are partitioned into primary and backup node sets, then the virtual nodes are mapped on substrate nodes of both sets, while the virtual links are mapped on disjoint primary and backup paths. Similarly, in [35], the authors proposed a parallel VN mapping algorithm, in which a modified Suurballe's algorithm is adopted to jointly optimize the mapping of primary and backup VNs. However, full backup provisioning for VNs is usually resource intensive and cost inefficient [38-43].

More commonly, the SVNE problem is solved by path protection, i.e., a virtual link is mapped on a primary substrate path and a disjoint backup path to protect against single-link failures. Literatures [33, 36-37] studied the dedicated path protection (DPP) SVNE schemes to combat against single substrate link failures. In [33], the authors employed a novel metric named as global resource capacity (GRC) for node mapping to balance the network load and improve the mapping acceptance ratio. While the authors in [36] and [37] utilized the Suurballe's algorithm for link mapping to minimize total length of primary and backup path pairs and optimize the VN embedding cost.

To optimize the backup bandwidth utilization, the authors in [38] and [34] proposed shared protection VN mapping schemes to combat against single-link failures. In [38], a dynamic cost model of sharable frequency slot (FS) was proposed to enhance intra-sharing in a VN and inter-sharing among VNs of the backup bandwidth. The authors in [34] designed a spare capacity assignment model to optimize backup bandwidth sharing among virtual links with the consideration of shared risk group (SRG), and employed a p-cycle based technique to re-optimize the backup bandwidth.

Considering the bandwidth resource efficiency of link mapping, the authors in [39, 40] employed the path splitting mechanism to split the virtual link into multiple sub-paths for better bandwidth utilization on small pieces of available spectrum. In order to obtain minimal backup redundancy in SVNE against single-link failures, authors in [41, 42] exploited path splitting to map a virtual link over multiple disjoint substrate paths as working paths and to preserve another disjoint substrate path as the backup path, which is denoted as partitioning dedicated path protection (PDPP). Since the demanded bandwidth is evenly distributed on the working paths and the bandwidth preserved on the backup path is equal to that of each working path, PDPP can be considered as a fixed path splitting scheme. For minimal resource consumption, the authors in [43] also employed path splitting to solve the problem of SVNE against multi-failures. In which, the number of backup paths and backup nodes are to be determined according to the demanded protection level and the SRG constraint.

Apparently, the SVNE problem is deduced from the VNE, some critical concerns in VNE are accordingly aligned with the objectives of SVNE, such as balanced substrate resource utilization, optimal embedding cost, and maximal VN acceptance ratio [44-55].

For node mapping sub-problem, the authors in [44] proposed maxmapping and minmapping schemes to achieve the balance of node resource utilization. In maxmapping, the virtual node with maximum demand of node resource is mapped on the substrate node with maximum available computing resource, which is exactly opposite to minmapping. Literatures [45-48] developed node mapping algorithms based on a node ranking approach to improve the VN acceptance ratio, which involves two kinds of parameters, i.e., capacity of the node resource and its total adjacent link bandwidth. However, this approach might select some substrate nodes with more node resource but with less bandwidth resource in some particular adjacent links. In addition, even if the bandwidth demand can be satisfied, it might increase the distance between the virtual nodes, causing high bandwidth consumption. In order to alleviate this weakness, the authors in [49-51] took more parameters into account besides the aforementioned two parameters, such as substrate nodal degree and distance metrics, so as to optimize the embedding cost.

For link mapping sub-problem, particularly in EONs, the authors in [52], proposed First Fit (FF) and Link List (LL) heuristics to process spectrum assignment. The FF maps virtual links on the first fit subcarriers, while the LL preferentially maps the virtual link with highest bit rate demand on the physical path with maximum available resource. In [53], the authors proposed a layered-auxiliary-graph (LAG) to solve VNE problems in EONs. By decomposing the physical infrastructure into layers, LAG converts the link mapping sub-problem into an integrated routing and spectrum assignment (RSA) problem. However, in LAG, the modulation format and the bandwidth of the FSs have to be predetermined, leading to some sacrifice on spectrum utilization. In [54], the authors defined the fragmented block size to calculate the weight of a substrate link in LAG. Thus, a virtual link can be mapped on a least-cost substrate path with good performance on spectrum defragmentation. Similarly, authors in [55] designed the fragmentation degree as link cost to find the best candidate path for link mapping, so as to improve the spectrum utilization.

## 2.3 Motivation

As SVNE problem is nondeterministic polynomial (NP)-hard, most of the existing literatures, such as [39-42, 44], separated it into node mapping and link mapping sub-problems and solved them independently. However, the lack of coordination between node mapping and link mapping might cause a high embedding cost. For example, in order to balance the nodal resource utilization, the adjacent virtual nodes might be mapped on distant substrate nodes, leading to high bandwidth consumption in link mapping [40]. Therefore, developing a simple and effective coordination between node mapping and link mapping is a significant concern [45-51].

For EONs, the available bandwidth capacity of each substrate link depends on the continuous and contiguous available FSs, as well as the modulation formats employed by these FSs [9]. Consequently, for most of the conventional node ranking approaches [45-48], it is difficult to accurately evaluate the total adjacent link bandwidth of the nodes during node mapping on EONs. Hence, developing a flexible and effective node mapping strategy to coordinate with link mapping is still an open challenge.

Since most of the VN request blockings occur due to the lack of available bandwidth resource [56], how to improve the bandwidth utilization and balance the traffic load on SNs is a crucial issue. Although path splitting can break through the bandwidth limitation on single path [39, 40] and solve the SVNE problem with low backup redundancy [41-43], there are some concerns about the path splitting: 1) In principle, multi-path deployment consumes more bandwidth than single-path deployment; 2) The number of the splitting paths is restricted by the nodal degree of a substrate node; 3) The application of path splitting technique can also increase the guard band consumption due to the non-overlapping constraint in EONs [41, 42]; 4) Furthermore, path splitting can help to utilize small pieces of bandwidth but might produce new and smaller spectrum fragments. With the constraints of spectrum continuity, contiguity, and non-overlapping, the spectrum fragmentation problem is also an important issue of SVNE over EONs [9, 11].

In this chapter, we investigate SVNE problem in EONs with the goal of minimizing the VN request blocking ratio, while protecting the VNs against single substrate link failures. In order to pursue optimal resource utilization but avoiding high computing complexity, we propose a two-stage coordinated SVNE scheme, named as APSS, to solve the aforementioned concerns [57]. In which, a concise node mapping strategy is developed to balance the nodal resource and coordinate with the link mapping, an adaptive path splitting policy is proposed to reduce the backup redundancy in link mapping, and an agile FS windows (FSWs) choosing mechanism is designed to mitigate spectrum fragmentation in EONs.

## 2.4 Network Model and Problem Statement

### 2.4.1 Network Model

In this chapter, we model the SN as a planar and mesh EON, and denoted it as $G^s(N^s, L^s)$, where $N^s$ is the set of substrate nodes and $L^s$ is the set of bidirectional substrate links. We assume that a substrate node $n^s (\in N^s)$ has $c_{n^s}$ amount of available computing resource and a substrate link $l^s (\in L^s)$ has $f_{l^s}$ number of available FSs. Note, the $f_{l^s}$ can be converted to available bandwidth $b_{l^s}$ according to a given modulation format. The VN request is modeled as $G^v(N^v, L^v)$, where $N^v$ is the set of virtual nodes and $L^v$ is the set of bidirectional virtual links. We assume that a virtual node $n^v (\in N^v)$ demands $c_{n^v}$ amount of computing resource and a virtual link $l^v (\in L^v)$ demands $b_{l^v}$ amount of bandwidth.

### 2.4.2 Problem Statement

Here, we study the problem of SVNE over EONs. Assume that each VN request arrives independently and stay in the EON for a period of time, then departs. For a VN request, all virtual nodes should be mapped on distinct substrate nodes with adequate computing resource, while each virtual link connecting a virtual node pair should be mapped on at least two link-disjoint substrate paths to protect against single-link failures, and the spectrum allocated on these substrate paths must be subject to continuity, contiguity, and non-overlapping constraints. The objective of SVNE is to minimize the VN request blocking ratio while guaranteeing the full protection against single-link failures.

## 2.5 APSS Scheme

In this section, we propose a two-stage coordinated scheme, namely APSS, which consists of three parts: anchor node strategy, adaptive path splitting policy, and FSWs choosing mechanism. For a VN request, APSS first exploits the anchor node strategy in node mapping stage to find an appropriate location area for candidate substrate nodes, where the distances between the substrate nodes are taken into account to limit the distance spans of virtual links. In link mapping stage, APSS employs adaptive path splitting policy to solve the routing issue of the virtual links, in which, full protection against single-link failures with partial backup resource is supported. Then, a FSWs choosing mechanism is utilized to allocate FSs on the corresponding substrate paths while mitigating spectrum fragmentation. In the following, the three components of the proposed APSS scheme are specified respectively.

### 2.5.1 Anchor Node Strategy (ANS)

For node mapping stage, in order to balance the utilization of the substrate node resource, ANS finds a substrate node $n^s (\in N^s)$ with the highest capacity of available computing resource to be the anchor node $n_A^s$. Then, the virtual node $n^v (\in N^v)$ with the highest demand of computing resource is mapped on the anchor node, which is denoted by Eq. (2.1).

$$n^v \xrightarrow{\max(c_{n^v})} n_A^s \tag{2.1}$$

$$d(n_A^s, n^s) \leq H, \quad \forall n^s \in N^s \tag{2.2}$$

In Eq. (2.2), parameter $H$ is defined as the maximum distance between the anchor node and other candidate substrate nodes, so as to restrict the distance spans of the virtual links and reduce their bandwidth consumptions. $H$ can be simply counted in hops or kilometers. By satisfying parameter $H$, ANS finds the candidate substrate nodes for the remaining virtual nodes. Then, based on the principle of the virtual node with maximum node resource demand is mapped on the substrate node with maximum available node resource, viz. maxmapping, ANS maps the remaining virtual nodes one by one on the candidate substrate nodes. The node mapping process of ANS must be subject to the following constraints.

Eq. (2.3) restricts that a virtual node $n^v$ can only be mapped on a substrate node $n^s$ with enough available computing resource. In Eq. (2.4), $\varphi_{n^v, n^s}$ indicates whether a virtual node $n^v$ is mapped on a substrate node $n^s$. Eq. (2.5) ensures that a virtual node $n^v$ can be mapped on only one substrate node, and Eq. (2.6) guarantees that a substrate node $n^s$ can host at most one virtual node of the same VN request.

$$c_{n^v} \leq c_{n^s} \tag{2.3}$$

$$\varphi_{n^v, n^s} = \begin{cases} 1, & \text{if } n^v (\in N^v) \text{ is mapped on } n^s (\in N^s) \\ 0, & \text{otherwise} \end{cases} \tag{2.4}$$

$$\sum_{n^s \in N^s} \varphi_{n^v, n^s} = 1, \quad \forall n^v (\in N^v) \tag{2.5}$$

$$\sum_{n^v \in N^v} \varphi_{n^v, n^s} \leq 1, \quad \forall n^s (\in N^s) \tag{2.6}$$

## 2.5.2 Adaptive Path Splitting Policy (APS)

For link mapping stage, in order to reduce the backup redundancy while guaranteeing protection against single substrate link failures, APS splits a virtual link into at most $K-1$ working paths and provisions one backup path. Here, parameter $K$ is a given integer which can restrain the excessive consumption of link resource. Thus, assume that $P_{l^v} = \{p_{l^v}^k\}$ is a set of disjoint substrate paths of link $l^v$ and $|P_{l^v}| \leq K$, it includes $|P_{l^v}|-1$ working paths and one backup path. The link mapping process of APS must be subject to the following constraints.

Eq. (2.7) ensures that, the bandwidth $b_{p_{l^v}^k}$ demanded by path $p_{l^v}^k$ cannot exceed the available bandwidth $b_{l^s}$ on link $l^s$ which is traversed by $p_{l^v}^k$. Eq. (2.8) uses $\xi_{p_{l^v}^k, l^s}$ to indicate whether $p_{l^v}^k$ traverses $l^s (\in L^s)$ and Eq. (2.9) ensures the disjointness of the paths in set $P_{l^v}$.

$$b_{p_{l^v}^k} \leq b_{l^s} \tag{2.7}$$

$$\xi_{p_{l^v}^k, l^s} = \begin{cases} 1, & \text{if } p_{l^v}^k (\in P_{l^v}) \text{ traverses } l^s (\in L^s) \\ 0, & \text{otherwise} \end{cases} \tag{2.8}$$

$$\sum_{p_{l^v}^k \in P_{l^v}} \xi_{p_{l^v}^k, l^s} \leq 1, \quad \forall l^s (\in L^s) \tag{2.9}$$

$$\sum_{p_{l^v}^k \in P_{l^v}} b_{p_{l^v}^k} - \max(b_{p_{l^v}^k}) = b_{l^v}, \quad \forall l^v (\in L^v) \tag{2.10}$$

Eq. (2.10) indicates that the bandwidth assigned on each path might differ from one another, which depends on the maximum available bandwidth on the substrate path. And the term of $\max(b_{p_{l^v}^k})$ represents the backup bandwidth, which can guarantee the full protection of $l^v$ against single-link failures. Besides, the number of paths employed by a virtual link, viz. $|P_{l^v}|$, is less than or equal to $K$, which can help to reduce the routing blockings due to the limitation of substrate nodal degree and can also reduce the total link resource consumption to some degree.

### 2.5.3 FSWs Choosing Mechanism (FCM)

When a substrate path $p_{l^v}^k$ is employed for link $l^v$, an appropriate FSW along $p_{l^v}^k$ should be chosen according to the demanded bandwidth $b_{p_{l^v}^k}$. For $b_{p_{l^v}^k}$, the required number of FSs $f_r$ can be calculated by Eq. (2.11). In which, $c_{FS}$ (in Gbps) is the bandwidth capacity of a FS, $g$ is a guard band to guarantee non-overlapping between two adjacent channels, and the $f_r$ number of FSs must be contiguous on the spectrum and continuous along the substrate path. $c_{FS}$ can be obtained by Eq. (2.12), where $c_{FS}^f$ is the spectrum of each FS (e.g., 12.5 GHz) and $s_{m_f}$ is the spectral efficiency of a modulation format $m_f$. Although employment of higher level of modulation format can provide better spectral efficiency, it is restricted by the transmission reach $r_{m_f}$ of the modulation format. Thus, Eq. (2.13) specifies that the length of a substrate path $d_{p_{l^v}^k}$ must not exceed the transmission reach of the employed modulation format.

$$f_r = \left\lceil b_{p_{l^v}^k} / c_{FS} \right\rceil + g \quad (2.11)$$

$$c_{FS} = c_{FS}^f \cdot s_{m_f} \quad (2.12)$$

$$d_{p_{l^v}^k} \leq r_{m_f} \quad (2.13)$$

In order to improve the spectrum resource utilization, FCM is designed to choose the FSWs which can mitigate the spectrum fragmentation and save more contiguous and continuous FSs for upcoming VN requests. Assume that the index of FSs is ranged from 0 to $l$ and the $i^{\text{th}}$ FS is denoted as $FS_i$. Eq. (2.14) defines $C_{l^s}^{FS}(i)$ as the cost of $FS_i$ on substrate link $l^s$ which indicates whether the $i^{\text{th}}$ FS is available or not. Further, a FSW is defined as $f_r$ number of contiguous and continuous FSs along a substrate path. If $FS_i$ is the first FS of a FSW, this FSW is denoted as $FSW_i$. Thus, Eq. (15) defines $C_{p_{l^v}^k}^{FSW}(i, f_r)$ as the cost of $FSW_i$ along the path $p_{l^v}^k$, which is a function of $C_{l^s}^{FS}(i)$. According to Eq. (2.15), a least cost FSW can be chosen and assigned to the corresponding virtual link.

$$C_{l^s}^{FS}(i) = \begin{cases} 1, & \text{if } FS_i \text{ is available} \\ 0, & \text{otherwise} \end{cases} \quad (2.14)$$

$$C_{p_{l^v}^k}^{FSW}(i, f_r) = \begin{cases} \sum_{l_s \in p_{l^v}^k} \left[ C_{l^s}^{FS}(i+f_r) \right], & \text{if } i = 0 \\ \sum_{l_s \in p_{l^v}^k} \left[ C_{l^s}^{FS}(i+f_r) + C_{l^s}^{FS}(i-1) \right], & \text{if } 1 \leq i < I - f_r + 1 \\ \sum_{l_s \in p_{l^v}^k} \left[ C_{l^s}^{FS}(i-1) \right], & \text{if } i = I - f_r + 1 \end{cases} \quad (2.15)$$

In the following, Fig. 2.1 illustrates the cost calculation for FSWs. We assume that the index of FSs on the substrate path 1-2-3 is ranged from 0 to 11 and the required FSs $f_r$ equals 2. Thus, the candidate FSWs can be $FSW_0(FS_0, FS_1)$, $FSW_4(FS_4, FS_5)$, $FSW_8(FS_8, FS_9)$, $FSW_9(FS_9, FS_{10})$ and $FSW_{10}(FS_{10}, FS_{11})$. According to Eq. (15), the costs of these FSWs are calculated as 0, 1, 3, 4 and 2, respectively. Evidently, choosing the least cost FSW can effectively mitigate FS fragmentation, while the greater the cost of the selected FSW is, the worse the fragmentation situation becomes.

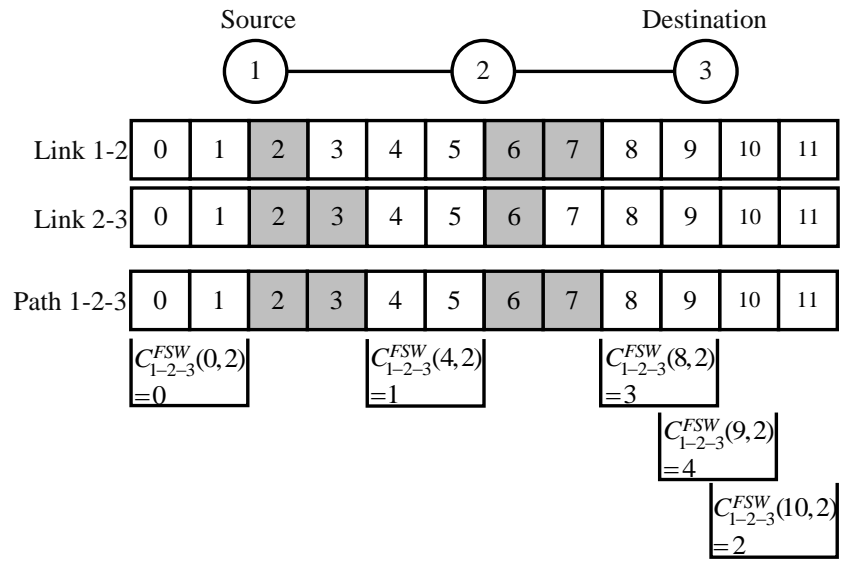

Fig. 2.1. Cost calculation for FSWs.

## 2.6. Heuristic Algorithm

Here, we denote VN mapping status as $M_{status}$, which is either SUCCESS or FAILURE. Similarly, $M_{status}^N$ and $M_{status}^L$ are denoted as node mapping status and link mapping status respectively, which are also either SUCCESS or FAILURE. For a substrate path $p_{l^v}^k$, the maximum number of continuous and contiguous FSs is denoted as $f_{p_{l^v}^k}^{\max}$.

### 2.6.1 Main-algorithm: Two-stage APSS Scheme

**Input:** $G^s(N^s, L^s)$, $G^v(N^v, L^v)$.

**Output:** $M_{status}$.

**Step 1:** Sort the virtual nodes of set $N^v$ and the virtual links of set $L^v$ in descending order of their demands for computing resource $c_{n^v}$ and bandwidth $b_{l^v}$ respectively. Sort the substrate nodes of set $N^s$ in descending order of their available computing resource $c_{n^s}$, let $N_A^s = N^s$.

**Step 2:** Take the first substrate node of $N_A^s$ as the anchor node $n_A^s$, execute Sub-algorithm 1 and obtain $M_{status}^N$, if $M_{status}^N = SUCCESS$, let $L_{temp}^v = L^v$, go to Step 3, otherwise, remove $n_A^s$ from $N_A^s$ and go to Step 4.

**Step 3:** For the first virtual link $l^v$ of $L_{temp}^v$, execute Sub-algorithm 2, obtain $M_{status}^L$, if $M_{status}^L = SUCCESS$, remove $l^v$ from $L_{temp}^v$, go to Step 5, otherwise, release all the substrate resources allocated to $G^v$, remove $n_A^s$ from $N_A^s$ and go to Step 4.

**Step 4:** If $|N_A^s| > 0$, go to Step 2, otherwise, mark $M_{status} = FAILURE$, go to Step 6.

**Step 5:** If $|L_{temp}^v| > 0$, go to Step 3, otherwise, mark $M_{status} = SUCCESS$, go to Step 6.

**Step 6:** Return $M_{status}$.

2.6.2 Sub-algorithm 1: Node Mapping Stage

**Input:** $G^s(N^s, L^s)$, $n_A^s$, $N^v$.

**Output:** $M_{status}^N$.

**Step 1:** Let $N_{temp}^v = N^v$ and put the anchor node $n_A^s$ and all of its adjacent (within a distance of $H$) nodes in set $N_{temp}^s$, if $|N_{temp}^s| \geq |N_{temp}^v|$, sort the elements of $N_{temp}^s$ in descending order of their available computing resource $c_{n^s}$, otherwise, mark $M_{status}^N = FAILURE$ and go to Step 6.

**Step 2:** Check the first virtual node $n^v$ of $N_{temp}^v$ and the first substrate node $n^s$ of $N_{temp}^s$, if $c_{n^v} \leq c_{n^s}$, go to Step 3, otherwise, mark $M_{status}^N = FAILURE$ and go to Step 5.

**Step 3:** Map $n^v$ on $n^s$, update $c_{n^s} = c_{n^s} - c_{n^v}$, mark $M_{status}^N = SUCCESS$, remove $n^v$ and $n^s$ from $N_{temp}^v$ and $N_{temp}^s$ respectively.

**Step 4:** If $|N_{temp}^v| > 0$, go to Step 2, otherwise go to Step 6.

**Step 5:** Remove all elements of $N_{temp}^v$ and $N_{temp}^s$, release all the substrate computing resources allocated to $G^v$.

**Step 6:** Return $M_{status}^N$.

### 2.6.3 Sub-algorithm 2: Link Mapping Stage

**Input:** $G^s(N^s, L^s)$, $l^v$, $K$

**Output:** $M_{status}^L$.

**Step 1:** Obtain the corresponding substrate node pair of virtual link $l^v$ and the demanded bandwidth $b_{l^v}$, let $k=1$.

**Step 2:** Calculate the $k^{th}$ disjoint shortest path $p_{l^v}^k$ between the substrate node pair, obtain $d_{p_{l^v}^k}$ and $f_{p_{l^v}^k}^{\max}$, calculate $c_{FS}$ and $f_r$ with demand of $b_{l^v}$ according to Eqs. (2.12) and (2.11).

**Step 3:** If $f_r \leq f_{p_{l^v}^k}^{\max}$, go to Step 4, otherwise, let $f_r = f_{p_{l^v}^k}^{\max}$, go to Step 4.

**Step 4:** Find the least-cost FSW on $p_{l^v}^k$ according to Eq. (2.15) and allocate it along $p_{l^v}^k$ as the working resource, update $b_{p_{l^v}^k} = c_{FS} \cdot (f_r - g)$ and $b_{l^v} = b_{l^v} - b_{p_{l^v}^k}$. If $b_{l^v} > 0$, go to Step 5, otherwise, go to Step 6.

**Step 5:** If $k < K-1$, $k = k+1$, go to Step 2, otherwise, mark $M_{status}^L = FAILURE$ and go to Step 8.

**Step 6:** $k = k+1$, calculate the $k^{th}$ disjoint shortest path $p_{l^v}^k$ between the substrate node pair, obtain $f_{p_{l^v}^k}^{\max}$, and calculate $c_{FS}$ and $f_r$ with demand of $\max(b_{p_{l^v}^k})$ according to Eqs. (2.12) and (2.11).

**Step 7:** If $f_r \leq f_{p_{l^v}^k}^{\max}$, find least-cost FSW according to Eq. (2.15) and reserve it along $p_{l^v}^k$ as the backup resource, mark $M_{status}^L = SUCCESS$, otherwise, mark $M_{status}^L = FAILURE$.

**Step 8:** Return $M_{status}^L$.

### 2.6.4 Compexity Analysis

According to the above procedure, we obtain the computing complexity of APSS, which is $O(K \cdot |L^v| \cdot |N^s|^2 \cdot \log|N^s|)$. Here, $|N^s| \cdot \log|N^s|$ is the complexity of shortest path algorithm, such as Dijkstra's.

### 2.7. Simulation and Analysis

In this section, we evaluate and analyze the performance of the APSS scheme, as well as its counterparts. In our simulation test, the parameter $H$ is set to 1 (hop), while the parameter $K$ is set to 4 for APSS. Here, we first introduce two baseline schemes into our simulation for comparison, i.e., MPF and MDF, according to [44, 42]. MPF employs maxmapping approach for node mapping and fixed path splitting approach for link mapping. The parameter $K$ for MPF is set to 3 or 4, in which one of the $K$ paths is deployed as backup path. Particularly, if $K$ equals 2, MPF turns to MDF, i.e., single path dedicated protection. Both MPF and MDF adopt First Fit approach for FS allocation. For highlighting each feature of APSS distinctly, we also introduce two more counterparts, i.e., APF and APC. Similar to MPF, APF employs fixed path splitting ($K = 3$ or $4$) for link mapping and First Fit approach for FS allocation, but utilizes anchor node strategy for node mapping. While APC employs FSWs choosing mechanism, fixed path splitting ($K = 3$ or $4$), and anchor node strategy.

## 2.7.1 Simulation Environment

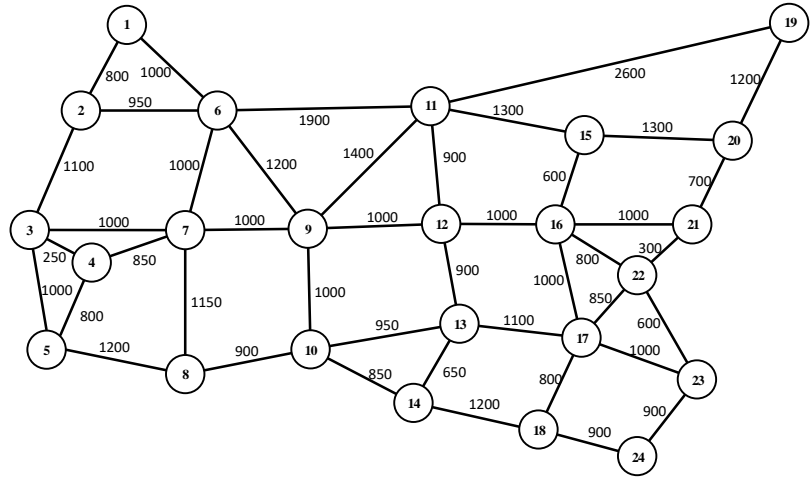

Fig. 2.2. USNET topology.

We employ USNET as the test SN. As shown in Fig. 2.2, the USNET has 24 nodes and 43 bidirectional links, and the numbers beside the links represent their distance spans in kilometers respectively. Assume that the computing capacity of each substrate node is 300 CPU units, while the spectrum capacity of each substrate link is the entire spectrum of C-band (4 THz). According to ITU-T G.694.1, we consider 12.5 GHz as the channel spacing, as well as the guard band spacing, thus each substrate link can accommodate 320 frequency slots in total. We employ four modulation formats, i.e., BPSK, QPSK, 8QAM, and 16QAM, and assume the corresponding spectral efficiency and transmission reach are (1, 2, 3, 4) bit/s/Hz and (4000, 2000, 1000, 500) km, respectively. For each VN request, the number of virtual nodes is randomly generated between 2 and 5, and the virtual link between a virtual node pair is generated with a probability of 0.5. Each virtual node demands a computing resource in the range of 7 to 10 CPU units, and each virtual link demands a bandwidth in the range of 25 Gbps to 250 Gbps.

We simulate a dynamic network environment, in which, the VN requests arrive in a Poisson process with arrival rate $\lambda$ and the average holding time is subject to a negative exponential distribution with mean value of $1/\mu$, thus the network load can be realized as $\lambda/\mu$ in Erlang. In each run of our simulation, 11000 VN requests are injected into the network. Considering the instability of network load at the initial stage, the first 1000 VN requests are not counted in our final results.

### 2.7.2 Numerical Analysis

In this chapter, four performance metrics are used to evaluate the SVNE schemes, i.e., average backup redundancy ratio (ABR), average guard band consumption (AGC), average spectrum consumption (ASC), and VN request blocking ratio (VBR). ABR is calculated as the number of FSs reserved for backup paths over the total number of FSs consumed by the accepted VNs. AGC is defined as the average number of FSs used as guard band for each accepted VNs. ASC is defined as the average number of FSs consumed by each accepted VNs. VBR is evaluated as the number of blocked VN requests over the total number of VN requests.

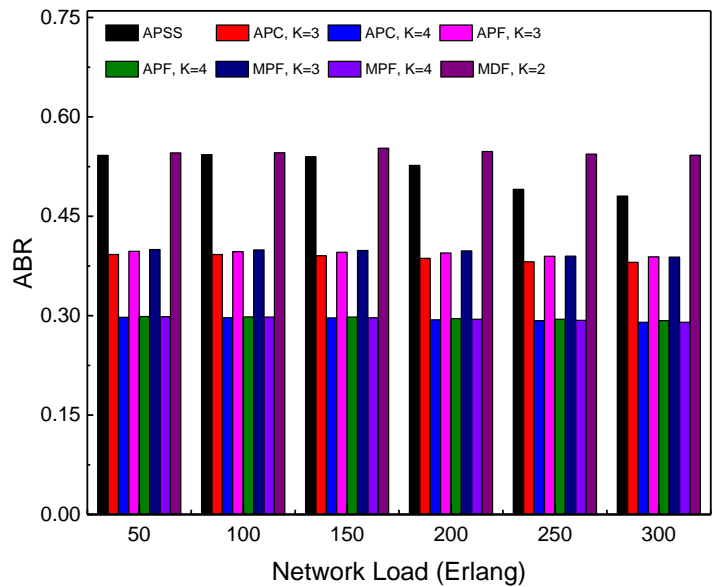

Fig. 2.3. Average backup redundancy ratio.

Fig. 2.3 shows the ABRs of the SVNE schemes, where, MDF as the primary baseline provisions a working path with demanded bandwidth and a dedicated backup path with the same amount of bandwidth for each virtual link, leading to the highest ABR, while MPF with $K=3$ significantly reduces ABR by evenly splitting the demanded bandwidth over two working paths and reserving 50% of the demanded bandwidth on a single backup path. And, as the value of $K$ increases from three to four, the bandwidth reserved on the backup path is reduced to about 33.3% of the demanded bandwidth, thus the ABR further decreases. We notice that APF and APC perform like MPF in terms of ABR due to the same path splitting policy, but the ABR of APSS is much higher than that of MPF, APF and APC, and decreases as the network load increases. This is because, APSS prefers to split the demanded bandwidth over as few working paths as possible, even if it can employ up to $K-1$ working paths with enough available contiguous and continuous FSs along the working path(s). Therefore, as shown in Fig. 2.4, the average number of working paths of APSS is close to one under low network load and gradually increases as the network load becomes high. It also illustrates that the spectrum fragmentation becomes worse under high network loads.

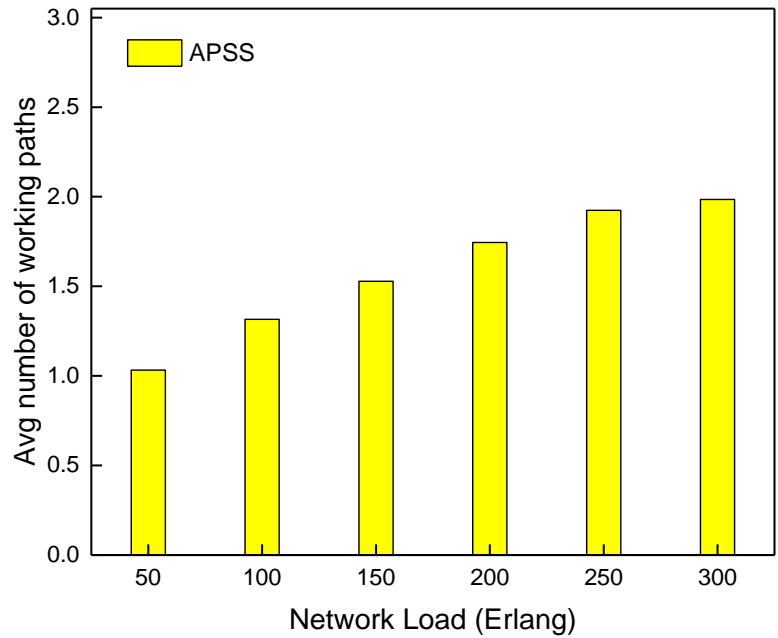

Fig. 2.4. Average number of working paths for APSS.

Fig. 2.5 presents the AGC performance comparison. We see that the bigger the value of $K$ is, the more guard band FSs are consumed. Compared with MPF, APF and APC consume less guard band FSs with the same $K$, because they utilize the anchor node strategy to restrict the candidate substrate nodes into a given area at node mapping stage, which can limit the hops of the substrate paths and save the guard band FSs at link mapping stage. So, as shown in Fig. 2.5, APF and APC with $K=3$ can even achieve a similar AGC performance as MDF for the same reason. Particularly, APSS outperforms all the others on AGC performance due to both anchor node strategy and adaptive path splitting policy. As the network load increases, APSS employs more paths to map the virtual link, therefore its AGC increases to some degree.

Numerical results in Figs. 2.3 and 2.5 indicate that path splitting can reduce the backup redundancy but consumes more guard band FSs. In Fig. 2.6, the ASC of MPF with $K=3$ is lower than that of MDF, while the ASC of MPF with $K=4$ is higher than that of MDF.

Which demonstrates that selecting an appropriate $K$ can help to gain a good tradeoff between backup redundancy and spectrum consumption. Compared with MPF and MDF, APF and APC can achieve better ASC performance because of its anchor node strategy. Furthermore, the ASC of APSS is lower than that of both APF and APC due to adaptive path splitting policy, and slightly increases when APSS employs more working paths under high network loads.

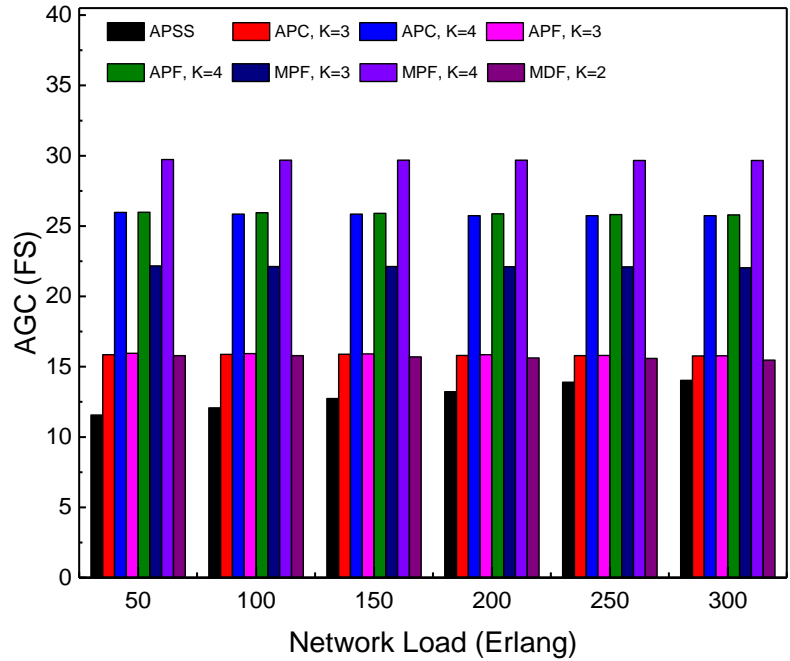

Fig. 2.5. Average guard band consumption.

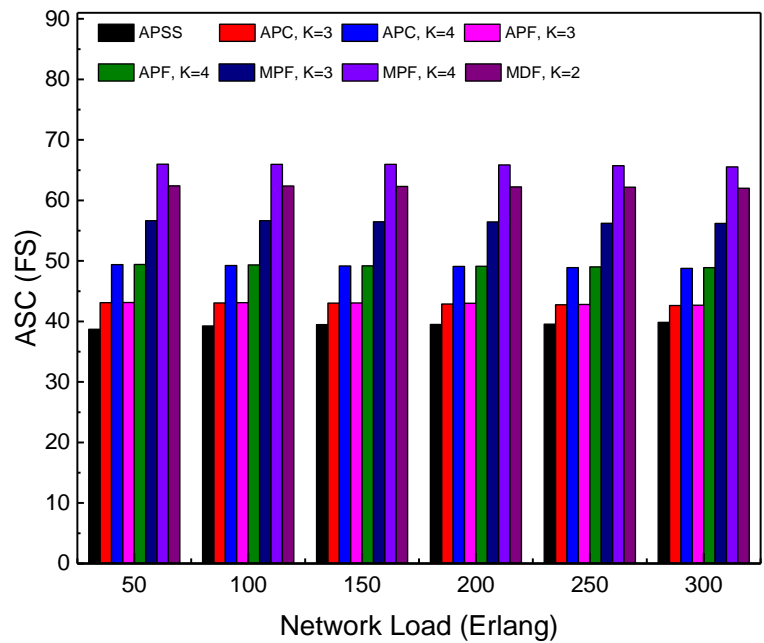

Fig. 2.6. Average spectrum consumption.

Fig. 2.7 illustrates the VBR performance, where APSS distinctly outperforms the other schemes. First, we notice that the VBRs of MDF and MPF with $K=4$ rapidly increase as the network load becomes high, which indicates that the available spectrum resource becomes scarce and the spectrum fragmentation gets worse under high network loads. In this case, it is difficult for MDF to find a big FSW along a single path to meet the full bandwidth demand, and it is also difficult for MPF with $K=4$ to find four disjoint paths for the required bandwidth, leading to high VBRs. We also see that the VBR of MPF with $K=3$ is much lower than that of MPF with $K=4$ particularly under high network loads. It demonstrates that evenly splitting the bandwidth over more paths can decrease the backup redundancy ratio, but also can consume more link resource and cause higher VBR. Compared with MPF, APF performs better on VBR performance because of its anchor node strategy. The VBR of APC is lower than that of APF, because the FSWs choosing mechanism in APC can mitigate the spectrum fragmentation problem to some degree. Furthermore, APSS adaptively splits the demanded bandwidth over working paths according to network resources, such as the availability of spectrum and the nodal degree. Thus, in practical, the average number of working paths of APSS is lower than that of APC, and the bandwidth of each working path can be different in APSS, which makes APSS more adaptable than APC, leading to better VBR performance.

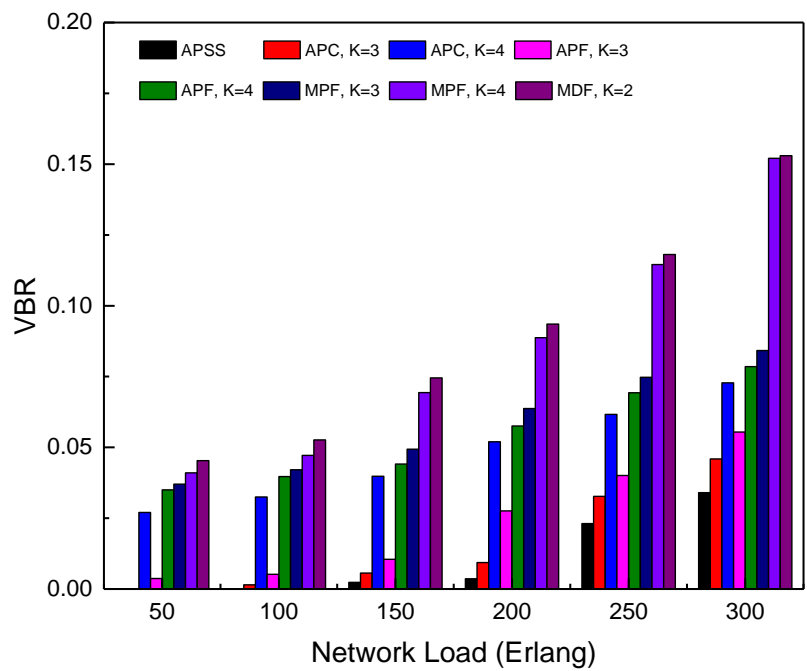

Fig. 2.7. VN request blocking ratio.

## 2.8. Chapter Conclusion

In this chapter, we investigated SVNE problem over EONs and proposed a two-stage coordinated SVNE scheme, named as APSS. The goal was to maximize the acceptance ratio of VN requests, while protecting against single-link failures. In order to balance the substrate node resource and optimize the link resource utilization, we proposed an anchor node strategy to narrow the solution space of node mapping so as to coordinate with link mapping, thus to limit the distance spans of virtual links. At link mapping stage, an adaptive path splitting policy was employed to reduce the backup redundancy, while taking the full advantage of available spectrum on the shorter paths. Furthermore, a FSWs choosing mechanism was designed to mitigate the spectrum fragmentation during FSs allocation. Numerical results illustrated that APSS, even if with higher backup redundancy ratio, can distinctly outperform the counterpart schemes in terms of VN request blocking ratio, average guard band consumption, and average spectrum consumption.

# Chapter 3 Synchronous Evacuation of VMs Under Disaster Risks

## 3.1 Chapter Overview

Network virtualization enables Internet service providers to flexibly provision heterogeneous VNs over shared SNs. As more and more applications are carried by VNs, the survivability of VNs becomes a critical challenge under large-scale disaster risks. In this chapter, we we study reactive VN survivability issue and propose a synchronous evacuation strategy for VNs with dual VMs inside a DRZ, which suffer higher risks than the VNs with single VM inside the DRZ. Aiming at the upcoming disaster destruction, the evacuation strategy first reconfigures the threatened VNs at outside of the DRZ, and the exploits post-copy technique to sustain the online service alive. We deploy the basic migration bandwidth for VM migration and upgrade the migration bandwidth after the VM downtime. For the migration process, the objective is to encourage parallel migrations and enhance bandwidth resource utilization. Then it enhances synchronicity between the dual-VM in order to shorten evacuation time. Numerical results show that the proposed strategy can achieve satisfactory performances in terms of average and total evacuation times of dual-VMs.

The rest of the chapter is organized as follows. In section 3.2, the background and related work are summarized. The motivation of this chapter is introduced in section 3.3. Section 3.4 presents the network model and problem statement. In section 3.5, the detailed synchronous evacuation strategy for dual VMs (SEDV) scheme is proposed and the corresponding heuristic algorithms are developed in section 3.6. Section 3.7 presents the simulation results and analysis. Finally, section 3.8 summarizes this chapter.

## 3.2 Background and Related Work

In recent years, the IT and telecom industries are evolving towards an era of network virtualization, which enables Internet service providers to flexibly provision heterogeneous VNsover shared SNs [12]. A VN usually consists of virtual nodes and virtual links, in which, the virtual nodes are realized through VMs running on substrate nodes (servers), and the virtual links are mapped on one or several successive substrate links with customized bandwidth. As more and more applications are carried by VNs, the survivability of the VNs becomes a critical challenge under large-scale disaster (such as earthquake) risks.

With a severe disaster threat, the VNs should evacuate the threatened VMs out of the DRZ as soon as possible. The VN evacuation mainly involves two processes: the first is VN reconfiguration (remapping), and the second is VM migration. In order to alleviate the impact on the online services during the evacuation, VM live migration would be a preferable manner. There are two fundamental live migration techniques, viz., pre-copy and post-copy [17]. In pre-copy, the storage content and memory pages (including dirty pages) are copied iteratively to the destination server, then the CPU state, as well as the remaining dirty pages, are transferred and the destination VM is resumed. In post-copy, the CPU state is transferred and the VM is resumed at the destination server, then the memory pages and storage content are pushed to the destination VM. Most existing literatures studied live migration of single VM and few ones researched on multiple VMs live migration [17, 58-60]. The authors in [4] studied live VM migration and designed a migration framework by utilizing central base image repository, in which the destination server downloads the base image from the central repository to optimize the data duplication and the number of pre-copy iterations. But it is expensive and not convenient to manage central repository for wide area networks. The authors in [59] focused on quickly migrating the target VM, so that the released resource can be utilized by other VMs. A mechanism named as enlightened post-copy was proposed, in which the most recent and current memory pages are transferred to the destination VM along with the CPU state, so as to minimize the fault page generation at the destination. Then, the other memory pages are actively pushed to the destination. Literature [60] studied the live migration of multiple VMs and designed a geometric programming model to optimize the bit rates for multiple VM migrations. The objective is to minimize the migration time of each VM which is defined as a tradeoff cost function with respect to the downtime (the duration of CPU state transfer) and the pre-copy time. However, as the migration process of each VM is still independent, the difference among the migration times of multiple VMs might influence the migration completion time for the individual VN.

## 3.3 Motivation

Most of the survivability works mentioned above mainly focused on providing protection and/or restoration for VN components with minimal resource consumption, but neglected the SLA constraint such as availability of VN applications and service interruptions. For example, the authors in [22] mentioned that their proposed protection approach can achieve higher resource efficiency and VN request acceptance ratio than its counterparts, but would cause more VM migrations and service interruptions. Whereas, the authors in [26] considered the periodical synchronization of VM images as a part of their protection scheme, so that the failed VM can be recovered from the previous saved state in the backup server. In [27], the authors took into account the continuous and acknowledged replication of VMs in different servers to guarantee zero VM state loss after a disaster failure. However, provisioning backup VMs at geographically separated servers and continuously replicating snapshots of VM's memory would consume huge amount of node resource and transmission bandwidth, leading to a high cost of application.

During a large-scale disaster (e.g., earthquake), the seismic waves propagating from the epicenter can produce immense damages over a wide area, which might cause geographically correlated substrate node destruction and/or link cuts, leading to a large amount of VN service disruptions and network resource crunch. Therefore, how to guarantee continuous VN service provisioning during a natural disaster is a critical challenge.

As VM live migration enable continuous service provisioning during the migration process, it can be implemented to sustain the online services alive during a disaster. Thus, migrating as many VMs as possible out of the DRZ as early as possible can be a feasible solution for VN disaster survivability. Also, the synchronicity among the VMs of the same VN can shorten the dual-VM evacuation time. Here, we study the VN survivability issue and design an evacuation strategy to tackle the synchronization problem for dual-VM migration under disaster risks [61].

## 3.4 Network Model and Problem Statement

### 3.4.1 Network Model

In this chapter, We model the SN as $G^s(V^s, E^s)$, where $V^s$ is the set of substrate nodes and $E^s$ is the set of substrate links. A substrate node $v^s(\in V^s)$ has certain amount of available node resources (e.g., computing, memory, storage) and a substrate link $e^s(\in E^s)$ has certain amount of available bandwidth capacity. A VN is modeled as $G^v(V^v, E^v)$, where $V^v$ is the set of virtual nodes and $E^v$ is the set of virtual links. A virtual node $v^v(\in V^v)$ is mapped on a substrate node $v^s(\in V^s)$ with required node resource and a virtual link $e^v(\in E^v)$ is mapped on a substrate path (consisting of one or several consecutive substrate links) with demanded bandwidth. For simplicity, in this paper, only one DRZ is assumed existing in $G^s$ at the same time, which covers two substrate nodes, as well as the adjacent links.

### 3.4.2 Problem Statement

Given a wide area network, we assume that there are two substrate nodes covered by a DRZ. In this case, the VNs with dual VMs inside the DRZ have higher destruction probability than those with single VM inside the DRZ. So, in this paper, we focus on the synchronous evacuation strategy for VNs with dual VMs inside a DRZ. The goal of this strategy is to minimize the evacuation time period for each dual-VM, as well as the entire time period for completing all dual-VM evacuations.

## 3.5 Synchronous Evacuation Strategy for Dual VMs (SEDV) Scheme

The proposed strategy, named as synchronous evacuation strategy for dual VMs (SEDV), mainly includes two phases, viz. VN reconfiguration and VM live migration.

### 3.5.1 VN Reconfiguration

For a VN $G^v$, the threatened dual virtual nodes and the corresponding virtual links are remapped outside the DRZ. The reconfiguration process follows two basic constraints. (i)

The candidate substrate nodes and links should have adequate available resource capacities to accommodate the remapped virtual nodes and links respectively. (ii) A virtual node can be mapped on only one substrate node, and a substrate node can host only one virtual node from the same VN. For optimizing the bandwidth resource, we narrow the solution space by limiting the distance between the candidate substrate nodes and the non-threatened virtual nodes of $G^v$, then choose the substrate nodes with minimum total distance for all the remapped virtual links of $G^v$.

### 3.5.2 VM Live Migration

The post-copy technique is employed to sustain the online services alive and minimize the total amount of data to be migrated. First, a basic migration bandwidth is set for all VMs to be migrated according to the link capacity of $G^s$, which can limit the maximum migration time for each VM. And a pair of shortest paths with the basic migration bandwidth is calculated for each dual-VM, which can prompt as many VNs as possible to evacuate at the same time. Then, the migration bandwidth of each VM is upgraded to the upper bound along its migration path and the migration is conducted. Here, we assume that $m_{i,j}$ and $m_{i,j'}$ are two threatened VMs of $G_i^v$. And the downtimes of $m_{i,j}$ and $m_{i,j'}$ are denoted as $t^d_{m_{i,j}}$ and $t^d_{m_{i,j'}}$, respectively. Thus, the downtime $t^d_{G_i^v}$ of $G_i^v$ is equal to $\max(t^d_{m_{i,j}}, t^d_{m_{i,j'}})$. When $G_i^v$ finishes its downtime $t^d_{G_i^v}$, the migration end times $t^e_{m_{i,j}}$ and $t^e_{m_{i,j'}}$ can be pre-calculated according to Eq. (3.1), where $D_{m_{i,j}}$ and $D^c_{m_{i,j}}$ are the total amount of data and the amount of data has been migrated at current time $t_c$.

When $t_c$ is equal to $t^d_{G_i^v}$, the migration bandwidths $b_{m_{i,j}}$ and $b_{m_{i,j'}}$ are adjusted according to Eq. (3.2) for synchronous migration.

$$t^e_{m_{i,j}} = \frac{D_{m_{i,j}} - D^c_{m_{i,j}}}{b_{m_{i,j}}} + t_c \qquad (3.1)$$

$$b_{m_{i,j}} = \begin{cases} b_{m_{i,j}}, & \text{if } t^e_{m_{i,j}} \geq t^e_{m_{i,j'}} \\ \dfrac{D_{m_{i,j}} - D^c_{m_{i,j}}}{D_{m_{i,j'}} - D^c_{m_{i,j'}}} b_{m_{i,j'}}, & \text{otherwise} \end{cases} \qquad (3.2)$$

## 3.6. Heuristic Algorithm

In the following, we develop our strategy into a heuristic algorithm. Assume that the set $S$ includes all of the threatened VNs, and the set $S_c$ includes the VNs under evacuation.

Step 1: Initialize $t_c = 0$, and set $S = \{G_i^v\}$.

Step 2: If $S \neq \emptyset$, for each $G_i^v \in S$, reconfigure the threatened virtual nodes and the corresponding virtual links, respectively pre-calculate the migration paths for $m_{i,j}, m_{i,j'} (\in G_i^v)$ according to the basic migration bandwidth and go to Step 3, otherwise, go to Step 4.

Step 3: If migration paths are successfully found for both $m_{i,j}$ and $m_{i,j'}$, upgrade $b_{m_{i,j}}$ and $b_{m_{i,j'}}$ to the upper bounds along the corresponding migration paths, move $G_i^v$ from $S$ to $S_c$, and execute the post-copy migration, go to Step 4, otherwise, go to Step 4.

Step 4: If $S_c \neq \emptyset$, go to Step 5, otherwise, Stop.

Step 5: For each $G_i^v \in S_c$, if $t_c = \min\{t_{G_i^v}^d\}$, calculate $t_{m_{i,j}}^e$ and $t_{m_{i,j'}}^e$ according to Eq. (1), adjust $b_{m_{i,j}}$ and $b_{m_{i,j'}}$ according to Eq. (2), go to Step 6, otherwise, go to Step 6.

Step 6: For each $G_i^v \in S_c$, if $t_c = \min\{t_{m_{i,j}}^e\}$, release the migration bandwidth resource assigned for $m_{i,j} (\in G_i^v)$, go to Step 7, otherwise, go to Step 2.

Step 7: If both $m_{i,j}$ and $m_{i,j'}$ of $G_i^v$ complete their migration, remove $G_i^v$ from $S_c$, go to Step 2, otherwise, go to Step 2.

## 3.7. Simulation and Analysis

We evaluate the performance of SEDV scheme and compare it with a baseline scheme named as best-effort evacuation for dual VMs (BEDV). BEDV employs the same reconfiguration scheme as SEDV, but has a different migration scheme. In the migration phase, BEDV calculates a pair of shortest migration paths and assigns full available bandwidth along them for each dual-VM. Two performance matrices are considered in this work, viz., total evacuation time (TET) and average evacuation time (AET). TET is defined as the entire time period taken to complete the evacuations for all threatened dual-VMs. AET is defined as the sum of evacuation time period for each dual-VM over the number of dual-VMs.

### 3.7.1 Simulation Environment

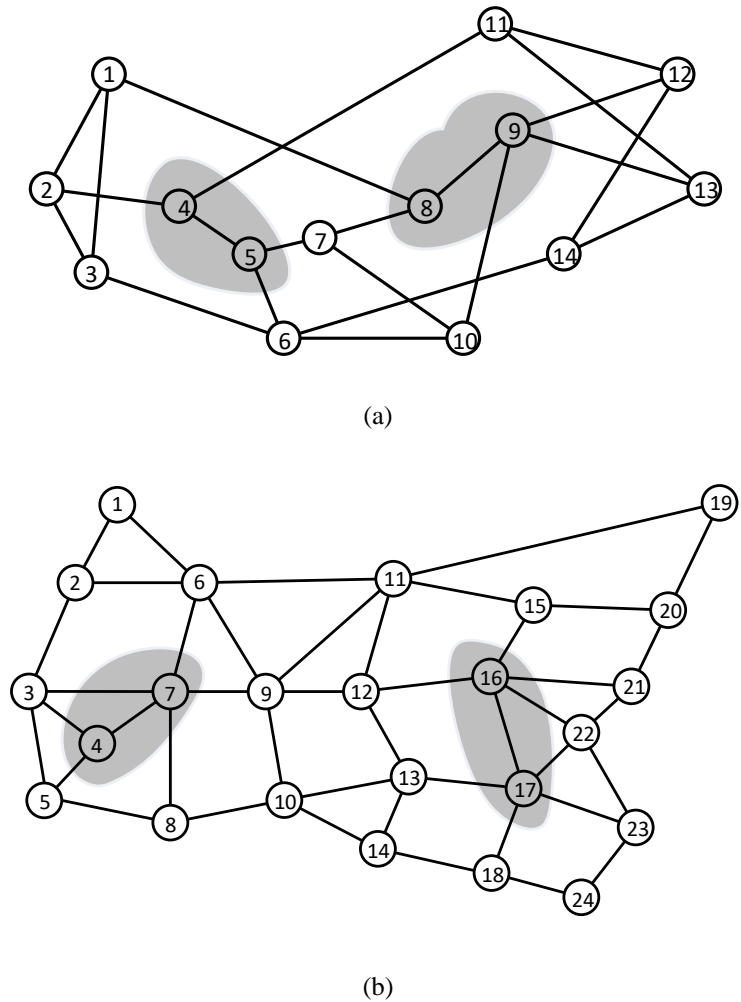

Fig. 3.1. Test Topologies (a) NSFNET (b) USNET.

As shown in Fig. 3.1, two test topologies, i.e., NSFNET, which has 14 nodes and 22 edges, and USNET which has 24 nodes and 43 edges, and the DRZs are marked with shadows. We assume that each substrate node has enough storage and computing resources, and each substrate link has the same amount of bandwidth capacity. In NSFNET, four different substrate link capacities are tested, including 65 Gbps, 70 Gbps, 75 Gbps, and 80

Gbps, and in USNET, four capacities are tested, including 130 Gbps, 140 Gbps, 150 Gbps, and 160 Gbps. Initially, there are 160 and 600 VNs randomly deployed in NSFNET and USNET, respectively. For each VN, the number of nodes is randomly generated between 3 and 5, and a virtual link is generated with a probability of 0.3 for each virtual node pair. The amount of data to be migrated for each threatened VM is uniformly distributed between 5 GB and 10 GB, and each virtual link demands an amount of bandwidth ranging from 0.5 to 3 Gbps. The downtime of a VM migration is randomly generated between 0.5 sec and 1.5 sec [62].

### 3.7.2 Numerical Analysis

We first evaluate the TET performance of SEDV with different basic migration bandwidths. As shown in Tab. 3.1 and Tab. 3.2, for different substrate link capacities, all TETs rapidly decrease as the basic migration bandwidth increases in the range of 0 to 4Gbps and becomes saturate after 5Gbps. Thus, we set the basic migration bandwidth as 5Gbps for the rest of the simulations.

Table 3.1. TET vs basic migration bandwidth in NSFNET.

| **Basic Migration Bandwidth (Gbps)** | **Link Capacity (Gbps)** | | | | |
|---|---|---|---|---|---|
| | 65 | 70 | 75 | 80 | |
| 0.1 | 106.711 | 50.5412 | 33.0738 | 25.4665 | **Total Evacuation Time (s)** |
| 1 | 39.8203 | 27.4123 | 19.1748 | 14.0703 | |
| 2 | 24.7524 | 19.0788 | 14.4859 | 11.7674 | |
| 3 | 23.3803 | 16.9475 | 14.5408 | 9.99412 | |
| 4 | 21.843 | 14.3593 | 12.4384 | 10.7167 | |
| 5 | 20.134 | 14.0694 | 11.5984 | 9.27406 | |
| 6 | 19.1494 | 14.2216 | 12.462 | 10.1627 | |
| 7 | 19.0595 | 14.3472 | 12.996 | 9.61849 | |
| 8 | 19.5225 | 12.9291 | 11.0584 | 9.76983 | |
| 9 | 19.2929 | 14.7511 | 12.179 | 10.7145 | |
| 10 | 19.2667 | 13.8156 | 12.5734 | 10.1457 | |

Table 3.2. TET vs basic migration bandwidthvin USNET.

| **Basic Migration Bandwidth (Gbps)** | **Link Capacity (Gbps)** | | | | |
| --- | --- | --- | --- | --- | --- |
| | 130 | 140 | 150 | 160 | |
| 0.1 | 46.3068 | 21.5263 | 15.3983 | 13.2301 | Total Time (s) |
| 1 | 17.9447 | 14.0857 | 9.02512 | 8.23799 | |
| 2 | 12.5627 | 9.51539 | 7.82105 | 6.86859 | |
| 3 | 9.79902 | 7.23926 | 6.9122 | 5.99374 | |
| 4 | 8.17874 | 7.09562 | 6.31106 | 5.74546 | |
| 5 | 8.13095 | 6.74518 | 6.33107 | 5.09895 | |
| 6 | 7.73132 | 6.51732 | 6.09131 | 5.15823 | |
| 7 | 7.33714 | 6.72401 | 5.17215 | 5.58146 | |
| 8 | 7.09122 | 6.30103 | 5.38614 | 5.3622 | |
| 9 | 6.79484 | 6.60195 | 5.29099 | 5.31627 | |
| 10 | 6.89984 | 5.71279 | 5.02908 | 5.03809 | |

Fig. 3.2 and 3.3 presents the AET performance in NSFNET and USNET, respectively, for both SEDV and BEDV. We see that SEDV distinctly outperforms BEDV, particularly under low substrate link capacity. The reason is that SEDV first guarantees an acceptable migration time for each VM by a basic migration bandwidth constraint and a bandwidth pre-upgradation, then guarantees a synchronous migration for the dual-VM by adjusting the migration bandwidth in proportion, so as to avoid the evacuation time of a dual-VM being prolonged by a single VM. However, BEDV finds a shortest migration path and occupies all available bandwidth along it for each VM, which might not guarantee an amount of migration bandwidth for keeping the migration time within an acceptable range particularly in the case of low substrate link capacity, as well as might cause severe imbalance for dual-VM migration, leading to a long evacuation time for the individual VN.

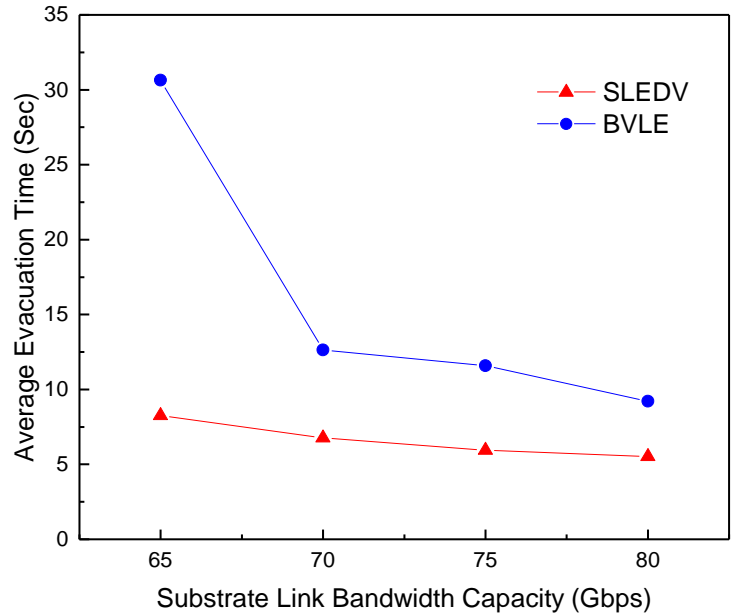

Fig. 3.2. Average evacuation time in NSFNET.

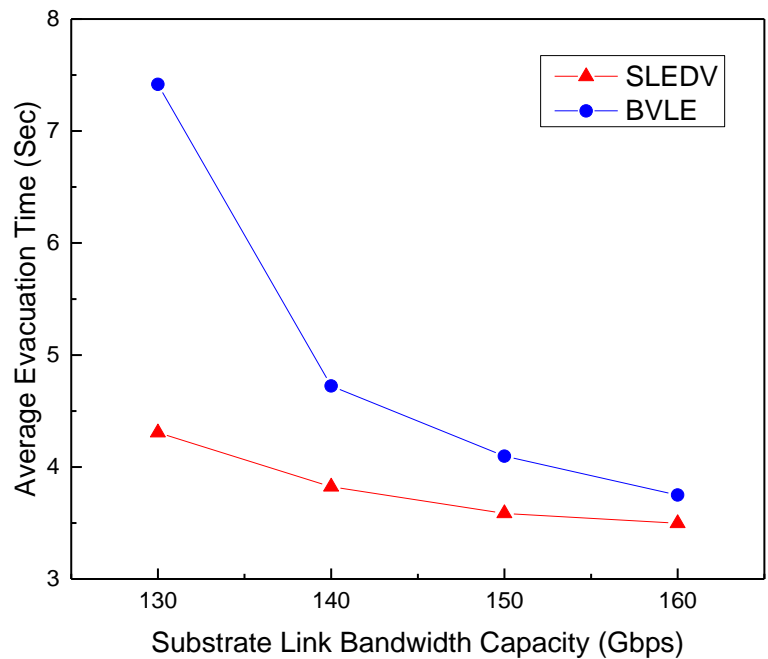

Fig. 3.3. Average evacuation time in USNET.

Fig. 3.4 and 3.5 shows TET performance for both of the schemes under different substrate link capacities, in NSFNET and USNET respectively. We see that the TETs of SEDV are much lower than that of BEDV. Because of the basic migration bandwidth constraint, SEDV prompts as many dual-VMs as possible to migrate at the same time. But, BEDV tries to assign maximum migration bandwidth for each VM, which practically cause a serial-style VM evacuation behavior. Coupled with imbalanced dual-VM migration, the TET performance of BEDV is worse than that of SEDV particularly under low substrate link capacity.

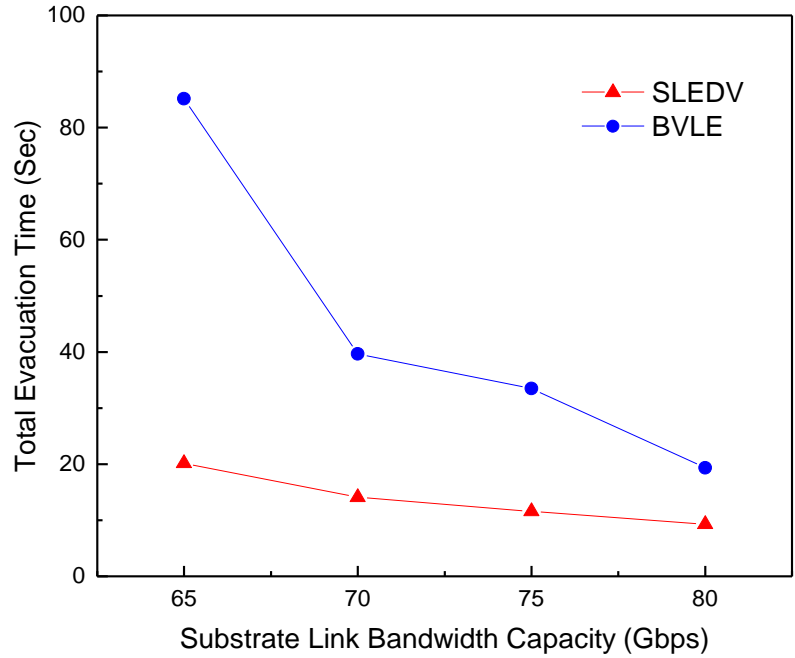

Fig. 3.4. Total evacuation time in NSFNET.

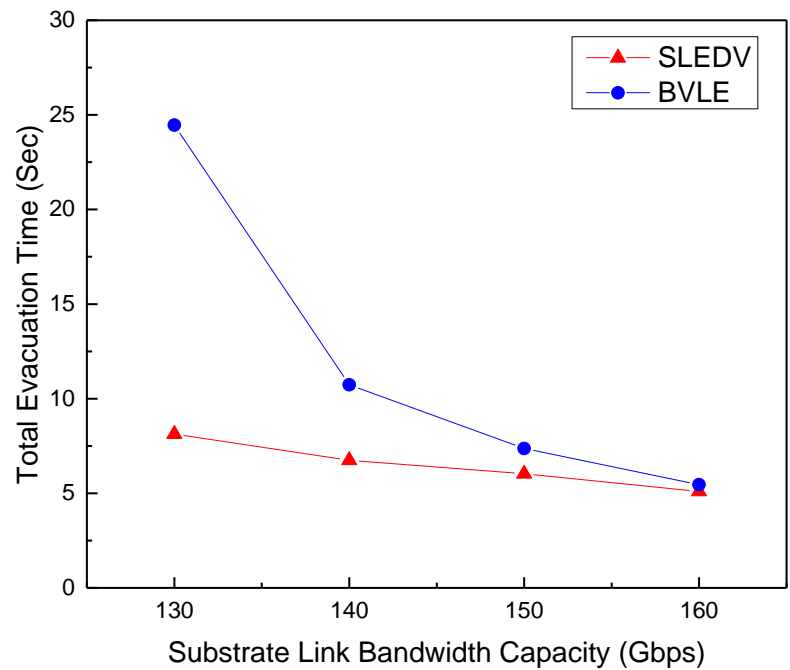

Fig. 3.5. Total evacuation time in USNET.

## 3.8. Chapter Conclusion

In this chapter, we studied VN survivability issue and concentrated on the synchronous evacuation problem for dual VMs under disaster risks. A synchronous evacuation strategy, named as SEDV, was proposed to evacuate the VNs with dual-VM within a DRZ as soon as possible. The SEDV first reconfigures the threatened VNs out of the DRZ, then utilizes post-copy technique, basic migration bandwidth constraint, and migration bandwidth adjustment to implement synchronous evacuation for the impacted dual-VMs. The simulation results illustrate that, under different network bandwidth resource cases, SEDV can gain satisfactory performance margin in terms of the average and total evacuation times compared with the baseline scheme.

# Chapter 4 Virtual Machine Migration Based Service Provisioning Strategy in MEC

## 4.1 Chapter Overview

In past decade, cloudification of the network applications centralizes the functions which can be provisioned to the end users connected to the network. As the future 5G, B5G applications thrust the strict QoS requirements, such as latency, cloud network cannot meet the QoS because of transmission and propagation delay in the longhaul network. To address this challenge, mobile edge computing (MEC) has become an effective solution which can meet the energy efficiency and delay requirement for the latency-constrained heterogeneous applications, such as vehicle to vehicle (V2V) communication, Industry 4.0, Internet of Things (IoT) applications for medical services, etc. [63]. In MEC systems, the end users can offload their tasks from the mobile devices to the nearest edge server consists of several VMs, which has the storage and computing capacity. In order to improve user experience, remote task offloading and redirection according the user mobility can be combined to accelerate the service migration [64]. In this chapter, we propose VM migration based fast service provisioning strategy in MEC systems to minimize the latency of the applications with strict QoS requirements. The proposed strategy traces the historic access patterns of the application and the basic image file of the VM-based application is located and fetched by the MEC server as soon as the request is submitted by the user. Then, the service starts processing at the MEC server. Along with that, to avoid the potential delay caused by incomplete migration, edge servers can continuously load the remaining pages in the image file. The edge servers repeat the same process if the mobile users change their location. As the service request starts it's processing as soon as the edge server fetches the basic image file of the VM, the potential latency for the service request is minimized. Furthermore, we replicate the basic image file and the modified pages of the VMs, to the nearest cloud server as backup and in future it can be cached to the edge servers to avoid the processing delay. Our main objective of this proposal is to minimize the latency to guarantee the QoS of the VM-based applications. This is an on-going research work and here we provide the system model and our designing methodology.

The rest of the chapter is organized as follows. Section 2.2 introduces MEC architecture and service provisioning in MEC. In section 4.3, we briefly discuss our future research's methodology, i.e., VM migration-based service provisioning strategy.

## 4.2 Introduction to MEC Architechture and Service Migration

The emergence of the IoTs has promoted rapid growth of lightweight mobile devices (for e.g., smart watches, medical IoT devices, body sensors, and vehicle equipments, etc.). However, due to the restricted storage, energy and computational capacities, IoT devices face significant challenges to fulfill the users' requirement need in latency-sensitive and context-aware applications. One feasible solution is to offload the data for processing from IoT devices to the cloud servers. Unfortunately, the uncontrollable transmission delay introduced by the Internet (basically by the longhaul network) can notably degrade the user experience. To address this challenge, we can bring the computing and storage capacities of the cloud servers at the edge of the network, which is near to the mobile devices. This paradigm is viewed as the MEC [65].

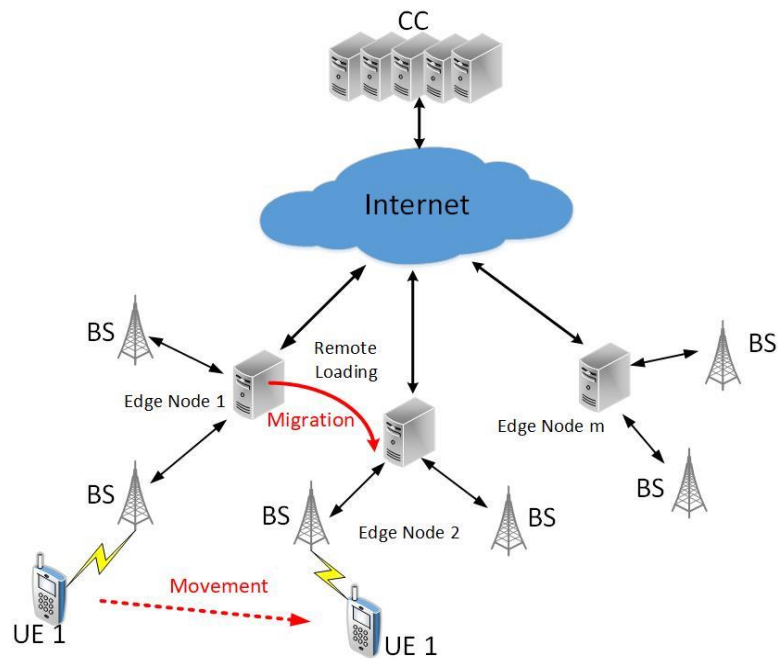

Fig. 4.1. Service migration in MEC.

In practice, as shown in Fig. 4.1, a MEC system consists of multiple edge nodes which are installed between wireless access networks and the core network to provide delay-sensitive services. From the system viewpoint, migrating workloads from clouds to edge nodes will significantly minimize the workloads of the core network, and enables IoT users to access computing resources at a lower latency. Also, from the user perspective, migrating the tasks (such as data processing) from IoT devices to nearest edge nodes can significantly reduce the energy requirements of mobile devices and boost the computation process [65].

Migrating the users' tasks to the MEC server is called as the task offloading. In MEC server, VMs have been utilized to isolate the service application logically, and to provide great flexibility with the utilization of the resources of MEC server. So, the service applications can be hosted by the particular VM located in the edge server. On the other hand, if the mobile devices are moving from one location to another location, such as vehicular devices, then its server-side service also required to be migrated from the original server to a new server closest to the current UE location, to ensure low latency. So, when the UEs offload their task to the MEC server, the server can directly start its computation if it holds the image files and backup files of the particular application. One service application can consist of hundreds of files, and several of these files overlap with other service application applications. So, duplicating all these files on each edge server or migrating the replicas to follow mobile devices is not desirable, as there are also limitations in storage and network capacity.

## 4.3 VM Migration-based Service Provisioning Strategy

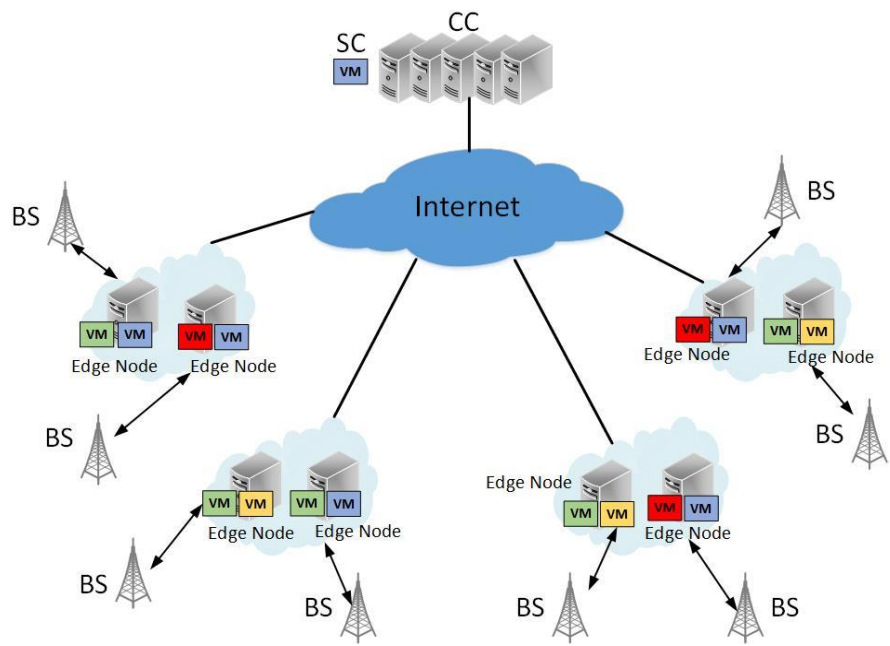

Fig. 4.2. VMs distribution in MEC system.

In order to minimize the latency of the service application, we propose a strategy which can boost up the service loading and execution time. This strategy follows the below process.

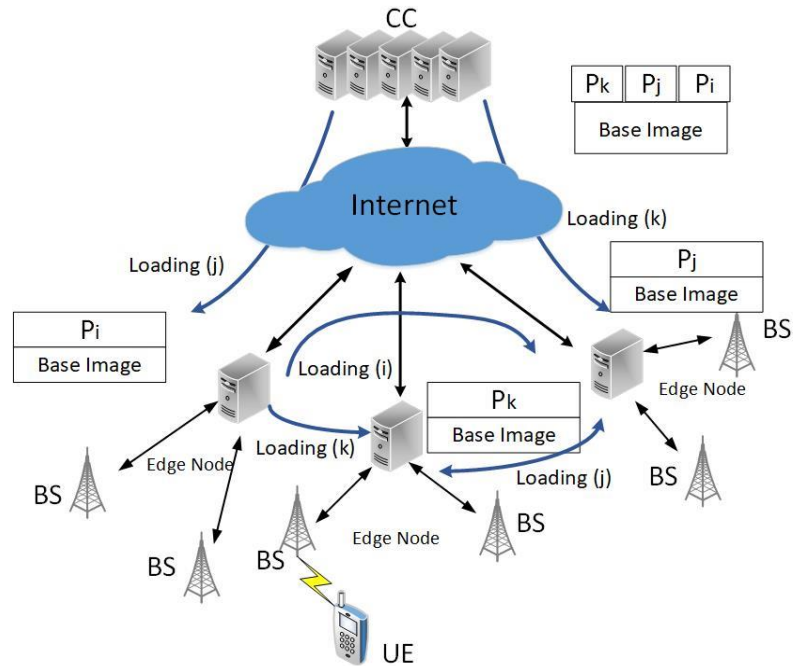

Fig. 4.3. Remote loading process and VM migration-based service provisioning.

- The proposed strategy first traces the historic access patterns of the application and distribute the VMs in MEC system. As shown in Fig. 4.2., different VMs are utilized for distinct service applications and they are distributed.
- As soon as the UEs offload their tasks to the nearest edge node, the edge server tries to locate the corresponding VM of the service application in the network.
- After that, the edge server fetches the basic VM image and starts excuting the task. If the user is not moving from one location to another location, then the whole task is executed on that particular edge server and delivered to the user.
- In another scenario, if the user is moving from one location to another location, then we exploit the post-copy VM migration strategy to support the user. Here, the original edge server (located at edge node) notifies the next server about the movement of the user. The later server then fetches the basic VM image from the corresponding VM of the application, which is similar as previous one. When the user reaches near to the new node, it offloads its remaining task to the new server. As the new server already has the basic image, it starts executing the task.

    Simultaneously, the previous server actively pushes the modified pages to the new server, which is named as remote loading and illustrated in Fig. 4.3. After completing the whole task, it is delivered to the user.

- Finally, the modified pages are replicated and cached to the original VM corresponding to the service application as the backup for future usage.

  So, this strategy has two-folded advantages. First, as the whole task is executed at the edge server near to the user, the transmission delay is minimized. Then, for executing the task the server is fetching the basic image file and starts execution. So, the execution time can also be mininmized. Simulatneously, the post-copy VM migration strategy is utilized for remote loading leading to reduce the total delay for delivering the service to the mobile users. This is an ongoing research. So, the formulations, detailed strategy, and the numerical results will be added in the later version.

# Chapter 5 Conclusion and future works

## 5.1 Conclusion

Network virtualization enables cloud service providers sharing various substrate resources and serving end users with heterogeneous VN applications. As more and more applications migrate to the cloud, the survivability of VNs has become a crucial issue. In the substrate network, even a single link failure can disrupt a lot of VN services. Particularly, during large-scale natural disasters (e.g., earthquake, tsunami, hurricane, etc.), some substrate nodes might be impacted or destroyed, even if the node equipment usually has higher anti-destruction grade than the long spanning cables. It would cause many VM failures and a large amount of VN service losses. In this dissertation, we investigated survivability issues of the VNs and proposed two schemes for two different disaster scenarios.

In first approach, we investigated SVNE problem over EONs and proposed a two-stage coordinated SVNE scheme, named as APSS. The goal was to maximize the acceptance ratio of VN requests, while protecting against single-link failures. In order to balance the substrate node resource and optimize the link resource utilization, we proposed an anchor node strategy to narrow the solution space of node mapping so as to coordinate with link mapping, thus to limit the distance spans of virtual links. At link mapping stage, an adaptive path splitting policy was employed to reduce the backup redundancy, while taking the full advantage of available spectrum on the shorter paths. Furthermore, a FSWs choosing mechanism was designed to mitigate the spectrum fragmentation during FSs allocation. Numerical results illustrated that APSS, even if with higher backup redundancy ratio, can distinctly outperform the counterpart schemes in terms of VN request blocking ratio, average guard band consumption, and average spectrum consumption.

In our second approach, we studied VN survivability issue and concentrated on the synchronous evacuation problem for dual VMs under disaster risks. A synchronous evacuation strategy, named as SEDV, was proposed to evacuate the VNs with dual-VM within a DRZ as soon as possible. The evacuation mainly includes two processes, viz., VN reconfiguration and VM live migration. For the threatened VNs, the evacuation strategy first re-maps them at the outside of the DRZ with minimal resource cost, then exploits post-copy technique to sustain the online service alive. During the VM migration process, the operations of basic bandwidth deployment and bandwidth upgradation are implemented, so as to encourage parallel migrations and maximize the resource utilization. After that, the migration bandwidth adjustment to implement synchronous evacuation for the impacted dual-VMs of the VNs. The simulation results illustrate that, under different network bandwidth resource cases, SEDV can gain satisfactory performance margin in terms of the average and total evacuation times compared with the baseline scheme.

## 5.2 The Future Works

In chapter 4, this dissertation also briefly introduced our future research work. As our future works, we are working on VM technology-based service migration strategy for MEC servers to minimize the processing delay and transmission latency for satisfying the QoS of the latency-constrained applications. In order to improve user experience, remote task offloading and redirection according the user mobility can be combined to accelerate the service migration. The proposed strategy traces the historic access patterns of the application and the basic image file of the VM-based application is located and fetched by the MEC server as soon as the request is submitted by the user. Then, the service starts processing at the MEC server. Along with that, to avoid the potential delay caused by incomplete migration, edge servers can continuously load the remaining pages in the image file. The numerical results and analysis will be published in later version.

# Acknowledgements

There are many people whom I wish to thank for their direct and indirect contributions in making this dissertation possible.

First and foremost, I want to thank my parents, Pradeep Sahoo and Rina Sahoo, and my aunty, Jharna Mahapatra, for supporting my dreams and ambitions in every possible way. Their encouragements have helped me to get through mental and intellectual tidal waves of the graduate studies.

Next, I would like to thank my advisor, Prof. Ning-Hai Bao, for the guidance, support, and honest criticisms he has provided throughout this work. I am grateful for the numerous meetings that he had with me to discuss our research problems and to guide me through them with his profound knowledge in many diverse areas. His mastery in the research domain has saved me on more than one occasion. Without his invaluable support and constant guidance, I could not have completed this dissertation.

Finally, I would like to thank my teachers from the coursework: Prof. Rong Chai, Prof. Taiping Cui, Prof. Xiaoge Huang and my friends: Sunny, Waleed, Abubakar, Ming Kuang, Liu Ziqian, Li Guo Ping, Yousuf, and Saifullah. Most importantly I am very much grateful to the Chinese government scholarship council for the generous scholarship to support my education in Chongqing, China.

# Publications and Achievements

**Academic Publications**

[1] Ning-Hai Bao, Subhadeep Sahoo, Ming Kuang, and Zhi-Zhong Zhang, "Adaptive Path Splitting Based Survivable Virtual Network Embedding in Elastic Optical Networks", https://doi.org/10.1016/j.yofte.2019.102084, Optical Fiber Technology, Nov., 2019.

[2] Ning-Hai Bao, Ming Kuang, Subhadeep Sahoo, Guo-Ping Li, and Zhi-Zhong Zhang, "Early Warning Time based Virtual Network Evacuation Against Disaster Threats", IEEE Internet of Things Journal. Dec., 2019, DOI- 10.1109/JIOT.2019.2963319.

[3] Ning-Hai Bao, Subhadeep Sahoo, Ming Kuang, and Zhi-Zhong Zhang, "Synchronous Evacuation Strategy for Double Virtual Machines Under Disaster Risk Zone", 3rd International Conference on Telecommunications and Signal Processing (TSP), Italy, Jul. 2020, DOI: 10.1109/TSP49548.2020.9163586.

[4] Subhadeep Sahoo, Xiao Han Dong, Zi Qian Liu, and Joydeep Sahoo, "Under Water Waste Cleaning by Mobile Edge Computing and Intelligent Image Processing Based Robotic Fish", Ericsson Innovation Awards 2019 (2nd Runner-up in North-East Asia Region).

**Patents**

"一种弹性光网络中基于锚点的虚拟网络抗毁映射方法 (Anchor Node Based Survivable Virtual Network Embedding Scheme in Elastic Optical Networks)", Application ID- 201910079410.2.

**Achievements**

[1] Regional Winner in North-East Asia Region in Ericsson Innovation Awards 2019.
[2] Awarded with prestigious CSC scholarship for pursuing Master of Science by Chinese government.
[3] Received Internship and Ph.D. offers from Nokia BELL Labs, Paris, France.
[4] Received Ph.D. offer in computer science from the University of California, Davis, USA.
[5] Received Ph.D. offer in computer science from the University College Dublin, Dublin, Ireland.
[6] Secured 1st position in article writing on "Building a community with shared future", organized by CQUPT.
[7] Received "Outstanding International Student" award by CQUPT.